\documentclass [12pt,twoside]{article}           
\usepackage[left,modulo]{lineno}
\usepackage{setspace}

\countdef\arxiv=201

\ifnum\arxiv=0 \input cpreb.tex \fi

\ifnum\arxiv=1

\usepackage{epsfig,times,lscape}
\usepackage[usenames]{color}

    \ifnum\count102=1

    \fi

    \ifnum\count102=2
\fi

\newcommand{\nc}{\newcommand}

     \ifnum\count101=1
\nc{\qI}[1]{\section{{#1}}}
\nc{\qA}[1]{\subsection{{#1}}}
\nc{\qun}[1]{\subsubsection{{#1}}}
\nc{\qa}[1]{\paragraph{{#1}}}

\def\qpar{\vskip 2mm plus 0.2mm minus 0.2mm}
\def\qL{\hfill \break}
     \fi 

      \ifnum\count101=2
 \nc{\qI}[1]{\parindent=0mm \vskip 8mm 
{\centerline{\LARGE \color{red}#1}}\vskip 3mm}
\nc{\qA}[1]{\vskip 2.5mm \noindent 
{{\bf\large\color{blue}  #1}} \vskip 1mm \parindent=0mm}
 \nc{\qun}[1]{\vskip 1mm \noindent {\sl #1 }\quad }

\def\qL{\hfill \break}
\def\qpar{\vskip 2mm plus 0.2mm minus 0.2mm}

      \fi

\def\qth{\vrule height 12pt depth 0pt width 0pt}
\def\qtb{\vrule height 0pt depth 5pt width 0pt}

\nc{\qfoot}[1]{\footnote{{#1}}}

      \ifnum\count101=1
\def\qbu{\hfill \par \hskip 6mm $ \bullet $ \hskip 2mm}
\def\qee#1{\hfill \par \hskip 6mm (#1) \hskip 2 mm}
      \fi
      \ifnum\count101=2
\def\qbu{\hfill \par \hskip 4mm $ \bullet $ \hskip 2mm}
\def\qee#1{\hfill \par \hskip 4mm (#1) \hskip 2 mm}
      \fi

\def\qparr{ \vskip 1.0mm plus 0.2mm minus 0.2mm \hangindent=10mm
\hangafter=1}

     \ifnum\count101=1 
 
     \fi
     \ifnum\count101=2

  \def\qcitb#1{\noindent \hbox to 102mm{\hfill \small #1} \vskip 1mm}
      \fi

 \def\qpages#1{\count102=0{\loop\advance\count102 by 1
 \null \vfill\eject \ifnum\count102<#1 \repeat}}

\def\qn#1{\eqno \hbox{(#1)}}

\def\qth{\vrule height 12pt depth 0pt width 0pt}
\def\qtb{\vrule height 0pt depth 5pt width 0pt}

\def\qv{\vskip 0.1mm plus 0.05mm minus 0.05mm}
\def\qhu{\hskip 0.6mm}
\def\qhv{\hskip 3mm}

\def\qhw{\hskip 1.5mm}
\def\qleg#1#2#3{\noindent {\bf \small #1\qhw}{\small #2\qhw}{\it \small #3}\qv }
\def\boxit#1#2{\vbox{\hrule\hbox{\vrule
\vbox spread#1{\vfil\hbox spread#1{\hfil#2\hfil}\vfil}%
\vrule}\hrule}}
\fi

\begin{document}
\thispagestyle{empty}



\markboth{{\sl \hfill  \hfill \protect\phantom{3}}}
        {{\protect\phantom{3}\sl \hfill  \hfill}}

\color{yellow} 
\hrule height 20mm depth 10mm width 170mm 
\color{black}
\vskip -1.8cm 

 \centerline{\bf \Large Deciphering infant mortality.  Part 1:
   empirical evidence}
\vskip 15mm
\centerline{\large 
Sylvie Berrut$ ^1 $,
Violette Pouillard$ ^2 $, Peter Richmond$ ^3 $ and Bertrand M. Roehner$ ^4 $
}

\vskip 4mm
\large

{\bf Abstract}\quad 
This paper is not (or at least not only) about human infant mortality.
In line with reliability theory, ``infant'' will refer here
to the time interval following
birth during which the mortality (or failure) rate decreases. 
This definition provides a
systems science perspective in which birth constitutes a 
sudden transition which falls within
the field of application of the {\it Transient Shock} (TS)
conjecture put forward in Richmond et al. (2016c). 
This conjecture provides predictions about
the timing and shape of the death rate peak.
(i) It says that there will be a death rate spike
whenever external conditions change abruptly and drastically.
(ii) It predicts that after a steep rising there will be
a much longer hyperbolic relaxation process. \qL
These predictions can be tested by considering
living organisms for which birth is a multi-step process.
Thus, for fish there are three states:
egg, yolk-sac phase, young adult. The TS conjecture 
predicts a mortality spike at the end of the yolk-sac phase,
and this timing is indeed confirmed by observation. \qL
Secondly, the
hyperbolic nature of the relaxation process can be tested
using high accuracy Swiss statistics which give postnatal death
rates from one hour after birth up to the age of 10 years.
It turns out that since the 19th century
despite a great overall reduction in infant mortality,
the shape of the age-specific death rate has remained
basically unchanged.
This hyperbolic pattern is not specific to humans.
It can also be found in small primates 
as recorded in the archives of
zoological  gardens.\qL
Our ultimate objective is to set up a chain of cases
which starts from simple systems
and then moves up step by step to more complex organisms.
The cases discussed here can be seen as 
initial landmarks.

\count101=0  \ifnum\count101=1
Whereas the process of aging has been (and still is)
much studied, the fast fall of the infant death rate in the hours, days,
weeks and months after birth has received fairly little attention.
The distinctive feature of this process is that the 
rate decreases as a power law with an exponent that is close to 1.
Understood in this sense, the phenomenon of
infant mortality exists not only in humans but also in animals and it
can even be found in technical systems. 
Our ambition is to set up a chain
that would start from the simplest systems and step by step
move up to more complex systems. \qL
In the present paper
we describe a number of landmarks on this road. For instance, we show
that in spite of the great reduction in overall human infant mortality
over the past century, 
the shape of the age-specific death rate has remained
unchanged. In fact, for the hours immediately after birth, the level
of mortality itself has hardly changed.\qL
Another key finding is the strong connection
between the degree of prematurity and the exponent of the death
rate fall.\qL
Regarding non-human cases, we introduce a 
distinction that we believe important
between living organisms (such as insects) which go through
different life stages and organisms (including plants)
which do not have several stages. Examples of both kinds are
shown and discussed. Only
the second display the typical power law fall. In 
our conclusion we emphasize that 
additional dedicated experiments are crucially needed.
\fi

\vskip 2mm
\centerline{\it \small Provisional. Version of 12 March 2016. 
Comments are welcome.}
\vskip 2mm

{\small Key-words: death rate, neonatal mortality, birth, primate, 
selection process, fish, yolk-sac.}
\vskip 2mm

{\normalsize 
1: Swiss Federal Office of Statistics, Neuch\^atel, Switzerland.
Email: Sylvie.Berrut@bfs.admin.ch \qL
2: Universit\'e Libre de Bruxelles (ULB), Brussels, Belgium
and University of Oxford (Wiener Anspach Foundation Postdoctoral
Fellowship).
Email: Violette.Pouillard@ulb.ac.be \qL
3: School of Physics, Trinity College Dublin, Ireland.
Email: peter\_richmond@ymail.com \qL
4: Institute for Theoretical and High Energy Physics (LPTHE),
University Pierre and Marie Curie, Paris, France. 
Email: roehner@lpthe.jussieu.fr
}

\vfill\eject

\large

\qI{Introduction}

There is a common saying according to which ``science starts
with the discovery of a pattern''. In the present paper,
we identify two patterns of infant mortality,
namely the yolk-sac pattern and the power-law death spike
pattern. Whereas the first pattern is fairly well understood,
the second raises questions for which we have no complete answer
so far. However, we believe that, taken together, these
patterns give us new insight into the mechanisms
of infant mortality.

\qA{Infant mortality seen as a transition death spike}

In a previous paper (Richmond et al. 2016c%
\qfoot{In a broader way, the present paper should be
seen as paper number 4 in a series of
socio- and bio-demography investigations 
started in Richmond et al. (2016a) and continued in
Richmond et al. (2016b,c).}%
)
it was suggested that any major
change in the conditions under which a system operates
will bring about an increased failure rate.
In the same paper this mechanism was shown to be at work
in the weeks following birth,
in the months following marriage and widowhood and also
when elderly persons are relocated in nursing homes.
In all these transitions between a state 1 and a state 2,
one observes a transient mortality spike
during which the death rate is temporarily multiplied by
a factor of 2 or 3 and even much more in the case
of the birth transition. 
\qpar

As explained in Richmond et al. (2016c),
the observed death spikes have a simple interpretation
as being a selection process. The systems which were 
adapted to state 1 but are not adapted to state 2 are
eliminated; this results in a death rate increase.
Although this explanation is satisfactory at a qualitative
level, at a quantitative level there remain
questions such as the following.
\qbu Can one establish a connection between the
characteristics of the transition on the one hand and
and the amplitude of the mortality spike on the other hand?
\qbu The infant mortality phase is characterized by
a death rate which decreases as a power law. Is this decrease
species dependent or does it  follow a general rule?
We will see that contrary to the increase 
of old-age death rates which are
very species dependent, the decrease of infant death rates
is fairly uniform: $ \mu=1/t{\gamma} $, where
the exponent $ \gamma $ is of the order of one.
\qpar

Most of the time expression the expression ``power law decrease''
refers to the slow decrease observed when
$ t \rightarrow \infty $. Here on the contrary, the
most conspicuous part of 
the mortality spike (Fig. 1a)
is the sharp decrease immediately following
$ t=0 $. Whereas a power law decrease is often meant
as a fall that is slower than an exponential, in the vicinity
of $ t=0 $ the fall is much faster than any exponential.
In order to emphasize this difference in what follows
we will use the expression ``hyperbolic power law''.
\qpar

Fig. 1a shows postnatal mortality in semi-log scale.
It gives a good idea of the shape of the spike.
%
\begin{figure}[htb]
\centerline{\psfig{width=8cm,figure=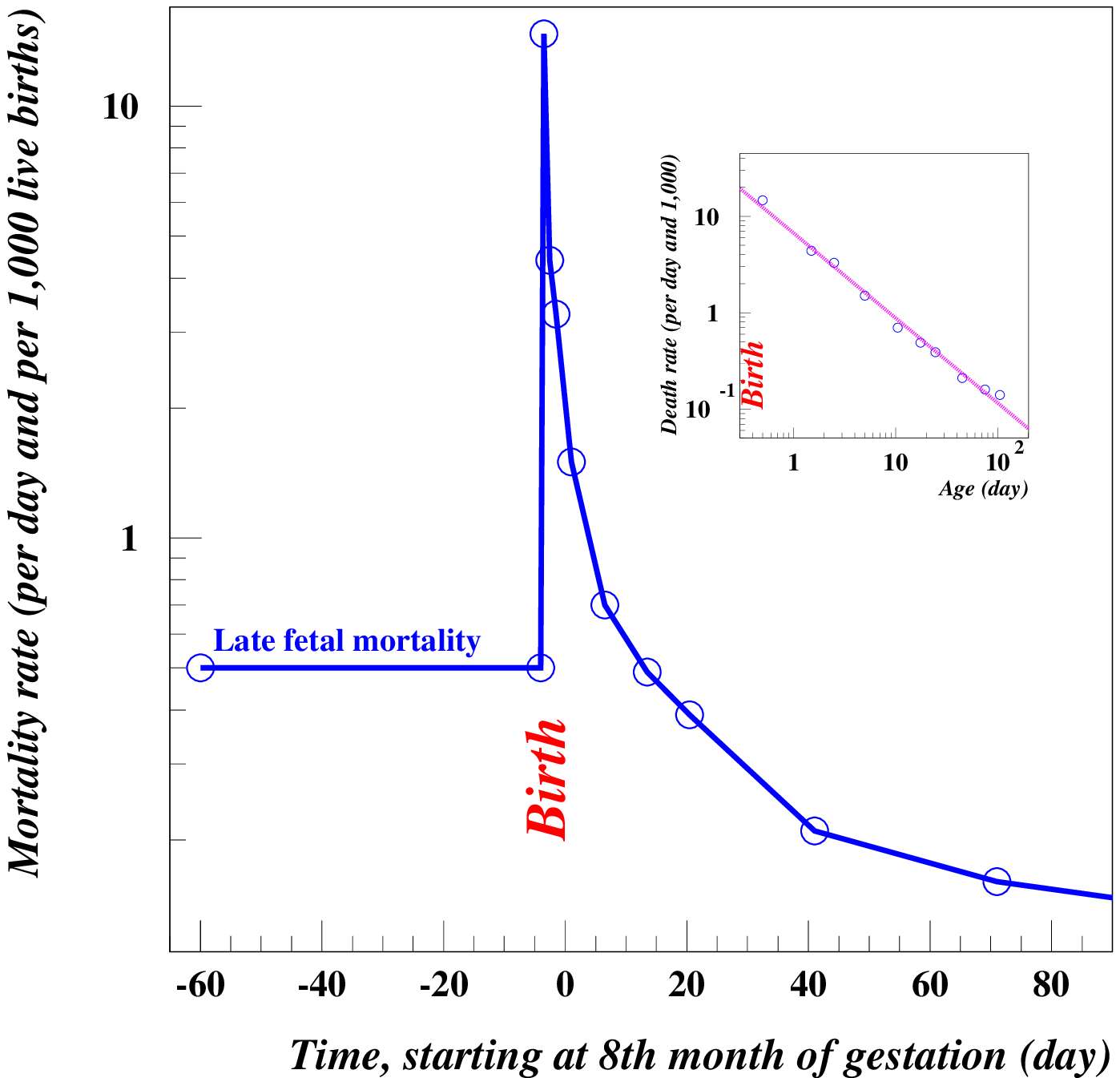}}
\qleg{Fig.\qhu 1a\qhv Transition from gestation to birth, USA 1923.}
{The graph shows the death rate spike which occurs after
birth.
The level section on the left-hand side schematically indicates
the (time-averaged) rate of late fetal mortality.
Then,
following birth, ``defects'' which were not of great consequence during
gestation suddenly lead to a dramatic increase
of the failure rate. The highest point corresponds to the first day,
the second and third points are for day 2 and 3 respectively. 
The fourth
point is the (daily) average for the age interval $ (3,7) $.
In the weeks and months following birth the death rate
decreases as a power law. For the inset log-log plot of the same data
the coefficient of linear correlation is $ 0.996 $
and the slope (i.e. the exponent of the power law) is 0.88.}
{Source: Linder and Grove (1947 p. 574-575).}
\end{figure}

However, in a semi-log representation one cannot identify 
the curve as a power law
and even more importantly one cannot explore a broad 
time interval. This is done in Fig. 1b.
%
\begin{figure}[htb]
\centerline{\psfig{width=12cm,figure=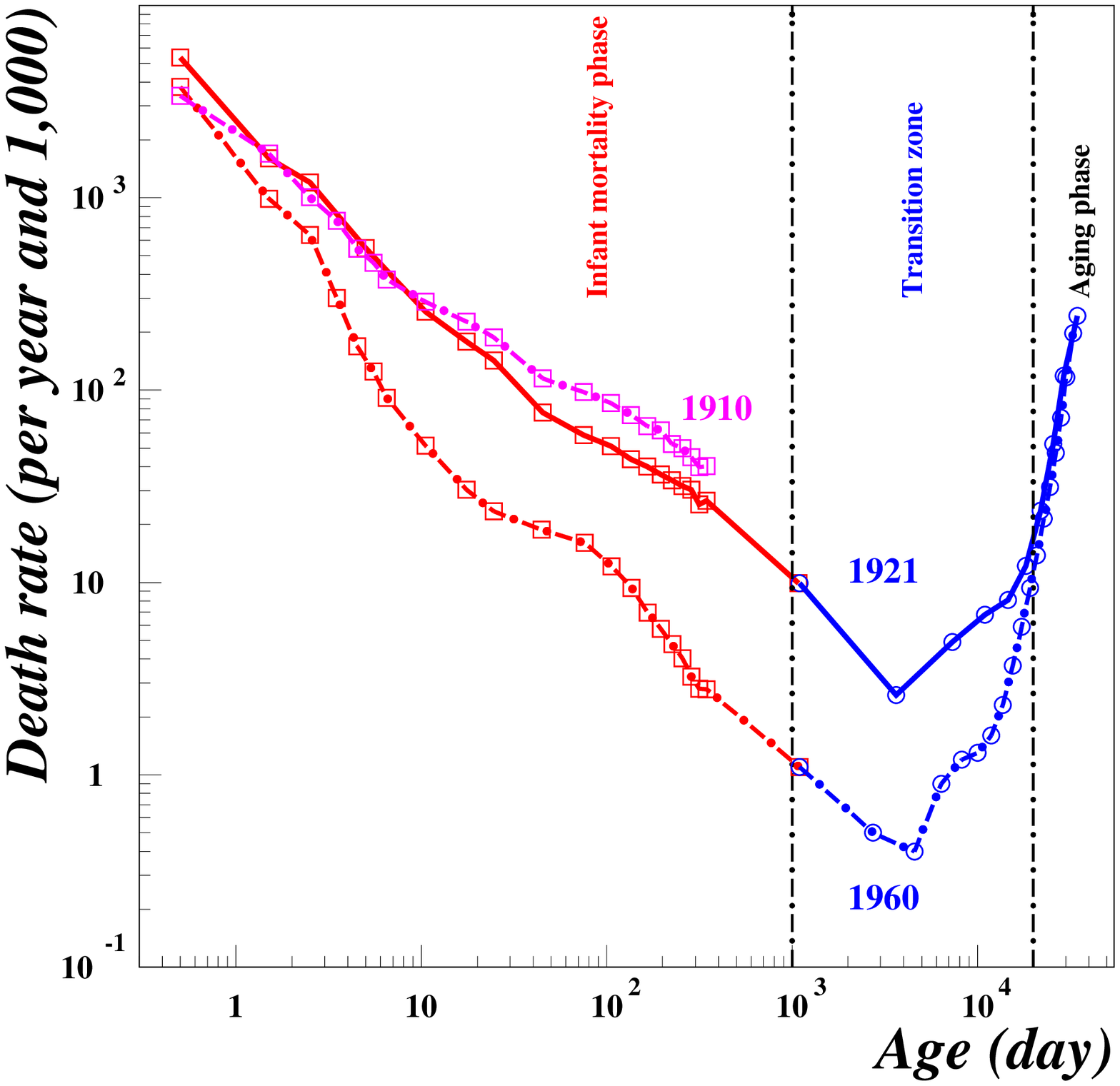}}
\qleg{Fig.\qhu 1b\qhv US death rates in log-log plot.}
{There are two phases (infant and aging) separated
by a transition. During the infant phase
the death rate decreases in an hyperbolic way. It can
be noted that despite medical progress the
two ends of the curves have not changed much.
In 1910 the slope of the falling regression line
(i.e. the exponent of the power law) was $ \gamma=0.65\pm 0.04 $
whereas in 1960 it was $ \gamma=1.01\pm 0.08 $.}
{Source: 1910: Mortality Statistics, Bulletin 109;
1960: Grove and Hetzel 1968, p. 210-211.}
\end{figure}
%
\qpar
It shows three phases: (i) The hyperbolic power law decrease.
(ii) A transition zone where the death rate is fluctuating
without any definite trend.
(iii) An aging phase where the death rate
increases sharply. The first and third phases are
certainly present in all living organisms. The first phase
is a selection process through which the items with
``manufacturing defects'' are eliminated. The increase
that occurs in the last stage is of course necessary if 
the death rate is to reach the level of 1,000 per 1,000
which signifies
total extinction of the cohort. Needless to say,
the respective length of each phase is species dependent.
As an illustration one can mention the case of
naked mole rates. We are told that these small mammals 
live until the age of 25 and that they remain in
good health until the very end of their life
(Buffenstein 2008).
This does not imply that the aging phase does not
exist but rather that it is short and therefore that the
increase of the death rate is very steep.
\qpar

\qA{Outline}

Physics relies on both experiments and theory.
The exploration of a new field usually
starts with a number of insightful
questions along with the experiments through which they
can be answered. This is what we try to do here.
Therefore, it should not come as a surprise that, at this
point, we do not propose a full fledged model.
At present the phenomenon of infant mortality 
is a
black box in the sense that we do not know
its mechanisms.
As a first step,
we need to identify the key variables.
Once we have got a better understanding it will be time
to express it in mathematical terms%
\qfoot{In the arXiv version of this paper two classes of
models are outlined. The first class proposes a connection
with the distribution of defects whereas the second establishes
a link between Gaussian and power-law random variables.}%
. 

The paper will proceed as follows.
\qee{1} In the next section we focus on an example
of infant mortality which is quite revealing because
it does not occur at birth (in this case hatching)
but at the end of the yolk-sac phase.
\qee{2} In the following section, we discuss the case of a simple
technical device, namely incandescent light bulbs. 
This will give us the opportunity to explain 
in a concrete way what should be meant by the expression
``lethal defect''. It will be seen that
in this case one can expect a
deterministic one-to-one correspondence between 
an initial defect that is present in a sample of items
and the age-specific failure rate. 
\qee{3} Then, we present human infant
mortality data.
The statistical evidence 
suggests that the death rate in the hours following birth
is largely independent of medical care.
In order to get a better insight into
the selection process at work during
the infant mortality phase we distinguish death rates
corresponding to various causes of death.
It will be seen that some follow a power law
behavior whereas others do not.
\qee{4} Next, in the hope of establishing a
chain of cases extending from ``simple '' to 
``complex'' systems, we present and discuss data for 
other species,
e.g. farm animals, primates, insects, plants.

\qA{Gompertz's law versus infant mortality}

Gompertz's law consists in the fact that in any human
population, after the age of 35, the death rate increases
exponentially with a doubling time of the order of 10 years.
For instance, in the United States in 1970, the death rate
in the age group $ 35-44 $ was 3.1 per 1,000 population of
both sexes, 7.3 for $ 45-54 $, 17 for $ 55-64 $,
36 for $ 65-74 $ and so on (Historical Statistics of the 
United States 1975, p. 60).
\qpar

Strictly speaking, infant mortality refers to the mortality
between birth and one year of age. However, we will use this
term in a broader sense for the whole period of time 
following birth during which the death rate decreases
before it levels off and starts to climb. For humans,
this phase extends from birth approximately to the age of 10 years.
In the present paper this phase will be referred 
to as the {\it infant phase}.
For humans (as well as for several other species)
it roughly coincides
with the period before sexual maturity. 
\qpar

In reliability studies, the time interval marked by a fall of the
failure rate is also called infant phase%
\qfoot{Other commonly used expressions are ``burn in phase''
or ``early failure period''.}%
.
The subsequent phase marked by an increasing failure rate is called
``wear out'' phase.

\qA{Birth transitions defined by the functions that must be switched on}

In order to survive an animal or a plant must be able to
use the oxygen contained in the air or in the water for
generating energy. It must also be able to find food and to
digest it. For animals finding food implies several challenges.
(i) Identification (ii) Moving to where it is located.
(iii) Swallowing and digesting.
In addition mammals
and birds need to regulate their body temperature.
Table 1 provides a summary of such functions in several 
types of species. 

\begin{table}[htb]

\small

\centerline{\bf Table 1 \quad Classification of transitions
according to the functions which are involved}

\vskip 5mm
\hrule
\vskip 0.7mm
\hrule
\vskip 2mm

$$ \matrix{
\hbox{Type}\hfill &  \hbox{State 1} \hfill &\quad& \hbox{State 2}\hfill
&\hbox{Oxygen} & \hbox{Food} &  \hbox{Food} & \hbox{Temperature}&
\hbox{Number} \cr
\qtb
\hbox{}\hfill &  \hbox{} &\quad& \hbox{}
&\hbox{} & \hbox{digesting} &  \hbox{finding} & \hbox{regulation}
& \hbox{of +}\cr
\noalign{\hrule}
\qth
\hbox{Mammal}\hfill &  \hbox{fetus} \hfill &\rightarrow&\hbox{newborn}\hfill
&\hbox{\LARGE +} & \hbox{\LARGE +} &  \hbox{\LARGE -} & \hbox{\LARGE
  +} & 3\cr
\hbox{Bird}\hfill &  \hbox{egg} \hfill &\rightarrow&\hbox{newborn}\hfill
&\hbox{\LARGE -} & \hbox{\normalsize +} &  \hbox{\LARGE -} &
\hbox{\LARGE +} & 1.5\cr
\hbox{Fish}\hfill &  \hbox{egg} \hfill &\rightarrow&\hbox{larva +
  yolk sac}\hfill
&\hbox{\LARGE -} & \hbox{\LARGE -} &  \hbox{\LARGE -} & \hbox{\LARGE
  -} & 0\cr
\hbox{Fish}\hfill &  \hbox{larva + yolk sac} \hfill &\rightarrow
&\hbox{larva (no yolk)}\hfill
&\hbox{\LARGE -} & \hbox{\LARGE -} &  \hbox{\LARGE +} & \hbox{\LARGE
  -} & 1\cr
\hbox{C. elegans}\hfill &  \hbox{egg} \hfill &\rightarrow&\hbox{larva}\hfill
&\hbox{\normalsize +} & \hbox{\normalsize +} &  \hbox{\LARGE +} &
 \hbox{\LARGE -} & 2\cr
\qtb
\hbox{Plant}\hfill &  \hbox{seed} \hfill &\rightarrow&\hbox{seedling}\hfill
&\hbox{\LARGE +} & \hbox{\normalsize +} &  \hbox{\LARGE +} &
\hbox{\LARGE -} & 2.5 \cr
\noalign{\hrule}
} $$
\vskip 1.5mm
\small
Notes: The {\LARGE -} sign means that the corresponding function
either was already ensured in state 1 or is unnecessary in state
2; e.g. temperature regulation is only necessary in homeothermic
species. On the contrary, the {\LARGE +} sign means that in order
to survive the newborn must be able to implement the
corresponding function. The {\normalsize +} sign indicates that the 
function was already ensured in state 1 but not exactly in the 
way necessary in state 2. For instance, bird embryos may be able
to digest the yolk contained in the egg but
unable to digest the food brought to them by
their parents.
The table suggests that the
birth of mammals involves more drastic changes than for the
other organisms mentioned in the table. Hence, one
expects a particularly high mortality spike.
Many other functions are of vital
importance (for fishes one can mention inflation of the swim bladder)
but most of them are in fact included in
the challenge of finding food because
this task requires to see (or smell), move, catch, swallow 
and digest.
\vskip 5mm
\hrule
\vskip 0.7mm
\hrule
\end{table}

For all the functions that need to be implemented in state 2
there is a non-zero likelihood of failure which will result in
an inflated death rate. In other words, one expects a mortality
spike each time a new function needs to be implemented.
This may happen at the time of birth but in some cases
it may also happen after birth. This is illustrated
in the next section.

\qI{The yolk-sac effect}

In the previous section we have seen that for the case
of human births the simultaneous
activation of 3 functions brings about a huge spike.
This naturally leads to question what happens when only one or
two functions are activated simultaneously. 
If this activation occurs
at a time $ t_1 $ after birth one would expect a
spike to occur around $ t_1 $. Does observation confirm
this prediction?
\qpar

What kind of organisms would be most appropriate for such a
test?
An idea which comes to mind is to use organisms
whose development goes through several stages.
This is the case of most insects. Worms giving
rise to flies or caterpillars giving rise to butterflies
are well known cases. From egg to adult
the development of
insects  involves several stages, not to speak of the
successive instars and moults.
However, for our purpose those stages are not 
really appropriate because they are too different from one another.
Feeding, for instance, is not at all the same problem
for a caterpillar and for a butterfly.
Fish larvae provide a better case.

\qA{Yolk-sac mortality spike for fish larvae}

When fish emerge from their egg they carry with them 
a yolk-sac which provides them with food until it is
exhausted. The insert of Fig. 2a shows such a yolk-sac
for salmons, a case in which it is particularly big.
Once the sac is depleted, the fish must find their
food themselves, a task which requires a whole chain
of functionalities: seeing, catching, swallowing and
digesting the food. If any of these functionalities
fails the fish will die from starvation within a few days. 
Thus, one expects a mortality spike in the days following
the end of the yolk-sac phase. 
\qpar
It is interesting to observe that
whereas this excess mortality has been commonly observed
it was often attributed to a number of special reasons (cannibalism,
unexpected changes in tank conditions) without 
real awareness of the underlying reason%
\qfoot{As illustrations one can mention the papers
by Lasker et al. (1970) and Garrido et al. (2015).}%
.

\qA{Anchovies and sardines}

In Fig. 2a the slope of the survival curve 
displays slight oscillations. However, given the scale
of the statistical fluctuations (shown by the thin lines)
it would be tempting to
discard them as being non significant. Yet, once death rates 
are computed a peak appears which coincides with the
end of the yolk sac phase as indeed shown by the starvation
curve.
\qpar
%
\begin{figure}[htb]
\centerline{\psfig{width=16cm,figure=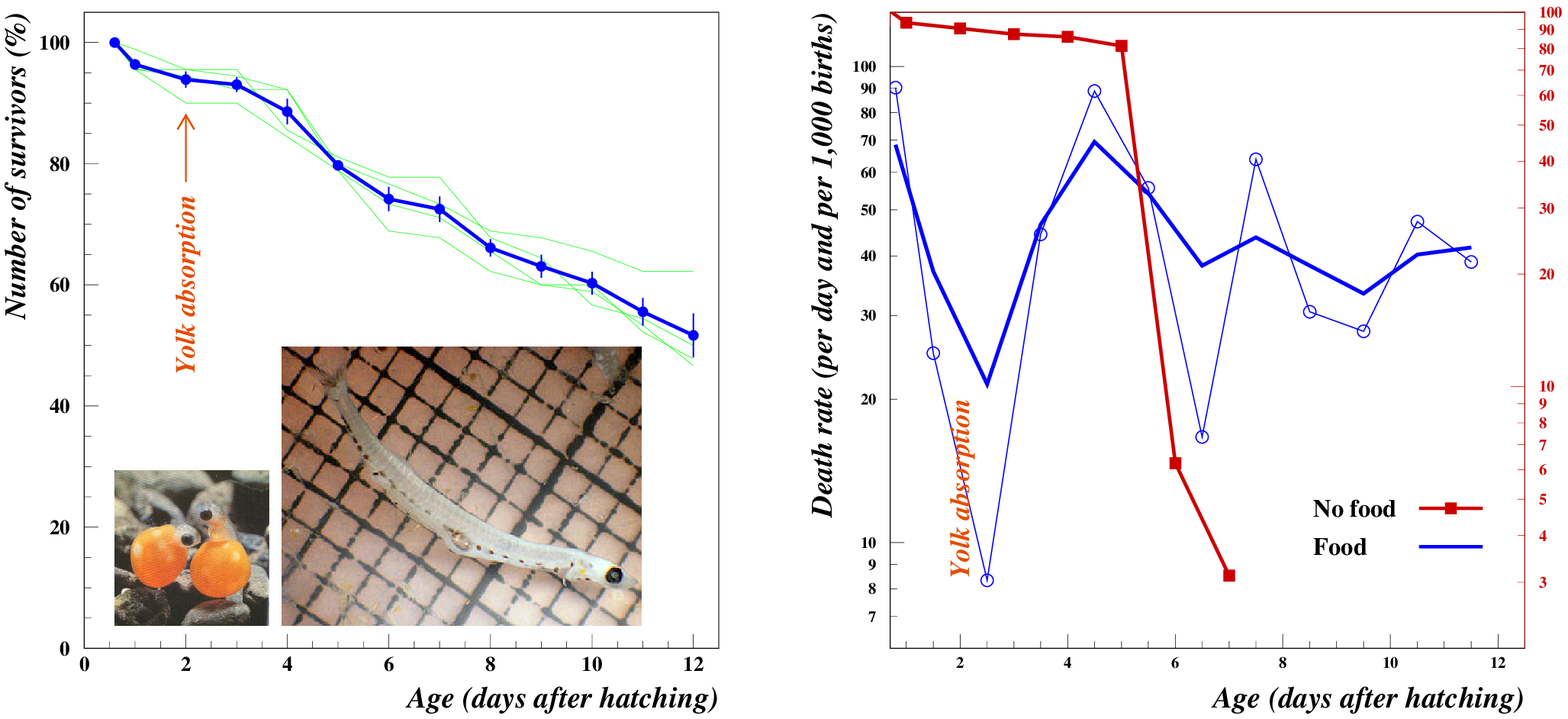}}
\qleg{Fig.\qhu 2a,b\qhv Yolk sac mortality spike for larvae of
California anchovy.}
{{\bf Left:} Survival curve of larvae of California anchovy
({\it Engraulis mordax}). The larvae were reared
in four 10 liter containers containing seawater and 100 eggs.
The thin (green) lines are the survival curves for the
four containers. The thick black line is their average $ m $
and the error bars indicate plus and minus the standard
deviation of the average.
The first insert shows two
larvae (not of E. mordax but of salmons which have
a larger amount of yolk)
shortly after birth
with their yolk sacs, while the second shows a larva
of E. mordax
whose length is 9mm which corresponds to an age of 18 days.
{\bf Right:} The thin lines display the experimental points.
The thick (blue) line is a 3-point centered moving average.
The peak between days 3 and 8 follows the depletion of the
food reserve contained in the yolk sac. 
The (red) line with the squares
corresponds to an experiment in which no
food was given; it shows that there is a time lag of about
3 days between food depletion and ensuing mortality.}
{Source: Lasker et al. 1970}
\end{figure}

A smilar effect can be found in the data published by
a team led by Susana Garrido (2015) who reared
820 larvae of European sardines ({\it Sardina pilchardus})
in laboratory conditions and recorded the number of 
deaths every day from hatching of the eggs to 60 days
later. For sardines the yolk sac phase lasts until they are
3 days old.
On the curve of the death rate as a function of age (not shown
here) there is
a decrease between $ t=0 $ and $ 5 $ days, followed by a sudden surge
in the interval $ (5,8) $ in which the death rate is multiplied
by 2. Then, the fall is resumed and continues until day 60.
In short, the mortality spike follows the depletion of the yolk sac
with a time lag of about 2 days.

\qA{Longer yolk-sac phases: redfish, sturgeons and salmons}

As the yolk-sac mechanism is common to all fishes which lay eggs,
many data should be available. Of particular
interest would be the larvae of salmons for in this case
the yolk sac stage lasts about 60 days. At that late age the
death spike should be particularly visible and clear.
For the present investigation one needs daily data for
populations reared in laboratory conditions.
We can offer the prediction that whenever appropriate
data become available the death rate of salmons should show
a peak around the age of 60 days. Let us hope 
the present paper may
encourage the publication of such data.
\qpar

In this subsection we present data for redfish (Fig. 3a,b)
and sturgeons (Fig. 4a,b).
In both cases the yolk-sac phase lasts about 12 
days. Redfish have the additional interest that, in contrast
with the majority of fish, the fertilization occurs
internally and the female spawns swimming larvae rather
than eggs. However, from our perspective this makes
little difference because the larvae carry a yolk-sac
just as when hatching occurs externally.
%
\begin{figure}[htb]
\centerline{\psfig{width=16cm,figure=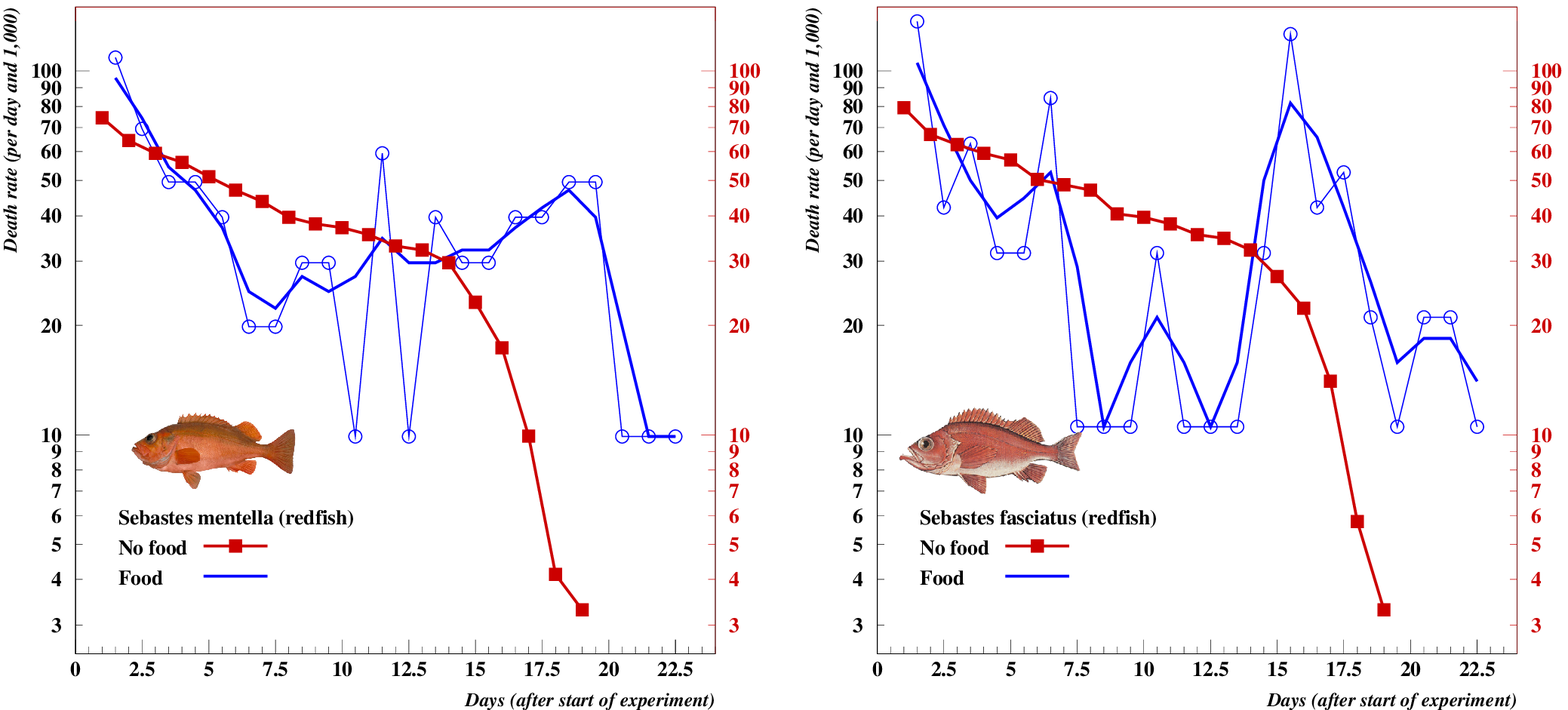}}
\qleg{Fig.\qhu 3a,b\qhv  Yolk mortality spike for two
species of redfish larvae}
{{\bf Left:} {\it Sebastes mentella };   
{\bf Right:} {\it Sebastes fasciatus.}
Each curve corresponds to a sample of 200 larvae.
The red curves with the squares show the percentages
of survivors when no food is given to them
even after their yolk-sac is depleted. In both graphs,
in accordance with expectation,
the mortality peak of nourished fish occurs in
synchronicity with the collapse of a population
that is not nourished.}
{Source: Laurel et al. 2001, p. 889}
\end{figure}

The data provided by Laurel et al. (2001) have a good 
side but also two drawbacks.
The good side is the fact that they include data for an unfed
group which allows us to know fairly exactly the moment
when the depletion of the yolk-sac becomes effective.
However, one drawback is the fact that the
small number of individuals in
each sample (namely 200 larvae) leads to fairly large
statistical fluctuations.
The second drawback is even more serious.
It consists in the fact that the larvae were collected offshore
and were ``stripped from ripe females'' collected by a
fishing ship. As not all females were
exactly at the same stage this collection method created large time
lags. This in turn led to fairly broad peaks extending over 
nearly 10 days. This effect is particularly obvious for
{\it Sebastes mentella.}
\qpar

What makes the data of Gisbert et al. (2000) of
particular interest is the fact that they rely on
samples of 2,500 larvae, i.e. ten times more
than in the previous experiment.
However, as this experiment (as well as all others)
was not designed for the purpose for which we are now
using it, there is also a downside, namely the fact
that there is no unfed sub-sample. This means that
one cannot identify exactly the end of the yolk-sac 
phase.

%
\begin{figure}[htb]
\centerline{\psfig{width=16cm,figure=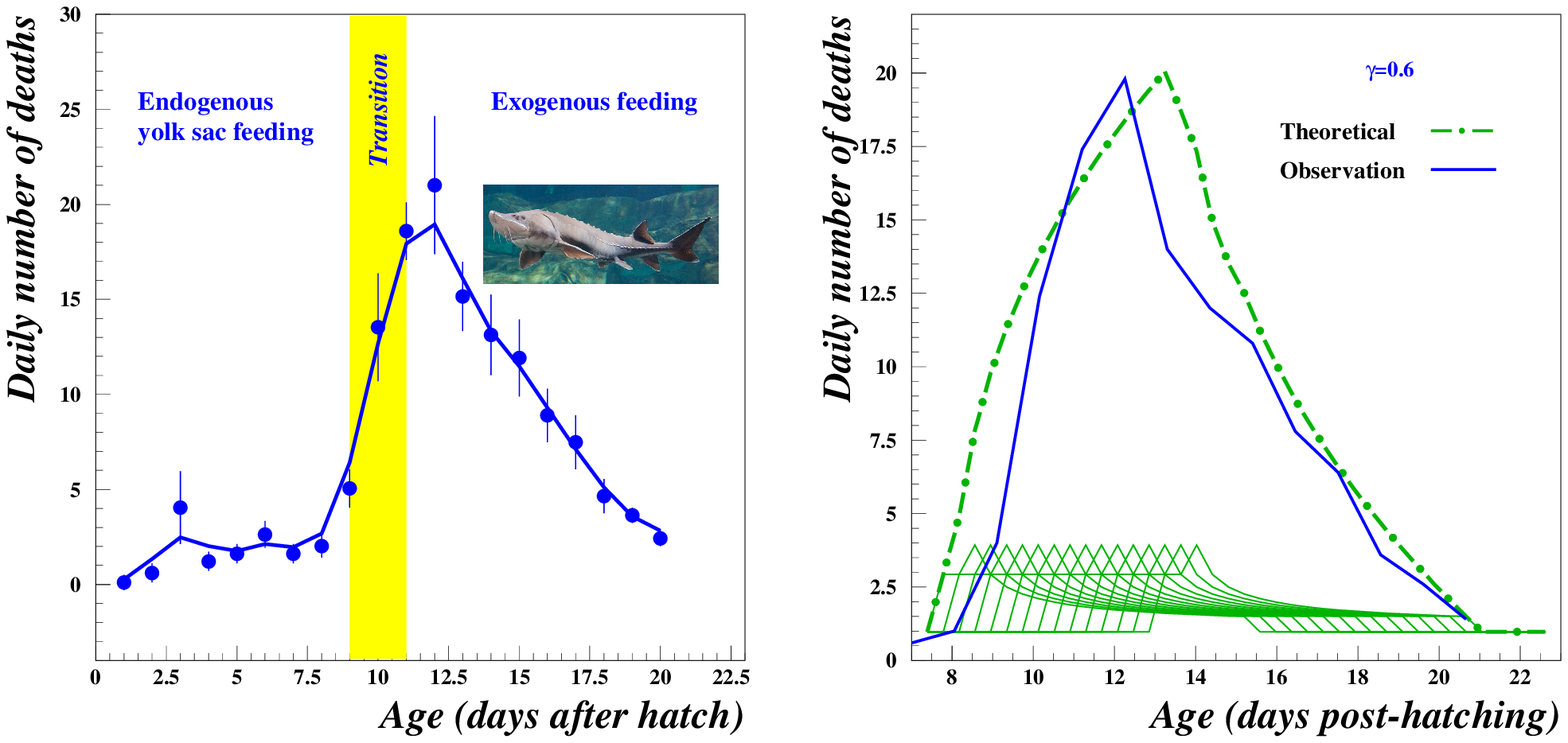}}
\qleg{Fig.\qhu 4a,b\qhv  Yolk mortality spike for Siberian
sturgeon larvae.}
{{\bf Left:} Each experiment involved 2,700 eggs. The duration
between spawning and hatching of the eggs was 7 days,
After
hatching the fish lived for 9-11 days on the food reserve
of their yolk sac. The transition to exogenous feeding was
marked by a major death peak. {\bf Right:} 
The fact that
the observed death rate
does not look like the spike of Fig. 1a can be due to
a dispersion in the age of
reaching the stage of exogenous feeding.
The curves
at the bottom (thin green lines) are hyperbolic death 
spikes of the form $ 1/(t-d)^{\gamma} $.
The (green) dot-dash line is their sum. The (blue)
solid line is the same curve as in (a).
The graph shows that the superposition of the
spikes 
produces a global curve that is fairly close to the
observed curve. For the sake of simplicity
we assumed a uniform dispersion of the time-lag $ d $;
the real (albeit unknown)
distribution of $ d $ is probably not uniform.}
{Source: Gisbert et al. 2000, p. 89.}
\end{figure}
\qpar
For sturgeons the yolk sac phase lasts about 10-12 days
which, although much shorter than the 60 days of salmons,
is substantially longer than in the cases of 
anchovy or sardines. As shown in Fig. 4a the transition
from endogenous to exogenous feeding is marked by
a major death peak. The fact that the transition
is distributed over 2 days results in a kind of moving
average; in other words, if all larvae were synchronized
the peak would likely be much sharper. 
This can be verified (Fig. 4b)
by superposing dispersed spikes and noting that the global
curve is fairly close to the observed mortality rate..
Incidentally, in the same paper the author
demonstrate that there is a correlation of 0.63 between
the diameter of the eggs and the 
time until first exogenous feeding. That property
can possibly be used to improve synchronicity.

\qA{Summary of noteworthy cases}

Table 2 summarizes the characteristics of cases
already explored 
in which the predictions based on the {\it Transient Shock}
conjecture were confirmed by observation. It gives also
two examples for which appropriate observations
have not yet been made (at least to our best
knowledge).

\begin{table}[htb]

\small

\centerline{\bf Table 2 \quad Testing predictions about
larvae mortality peaks.}

\vskip 5mm
\hrule
\vskip 0.7mm
\hrule
\vskip 2mm

$$ \matrix{
&\hbox{Species}\hfill &  \hbox{Number} \hfill & \hbox{Hatching}
& \hbox{End}\hfill
&\hbox{Predicted} \hfill & \hbox{Test of} \hfill & \hbox{Reference}\hfill
  \cr
&\hbox{}\hfill &  \hbox{of} \hfill & \hbox{rate}\hfill
& \hbox{of}\hfill
&\hbox{interval} \hfill & \hbox{prediction} \hfill & \hbox{of}\hfill 
\cr
&\hbox{}\hfill &  \hbox{organisms} \hfill & \hbox{}
& \hbox{yolk}\hfill
&\hbox{of death} \hfill & \hbox{through} \hfill & \hbox{data}\hfill 
\cr
&\hbox{}\hfill &  n & \hbox{}
& \hbox{phase}\hfill
&\hbox{rate spike} \hfill & \hbox{observation} \hfill & \hbox{}\hfill 
\cr
\qtb
&\hbox{}\hfill &  \hbox{} \hfill & \hbox{[\%]}
& \hbox{[day]}
&\hbox{[day/hour]}  & \hbox{} \hfill & \hbox{}\hfill 
\cr
\noalign{\hrule}
\qth
1&\hbox{California anchovy}\hfill & \hfill 400  & \hbox{?} & 2
& 4-7 & \hbox{C} & \hbox{Lasker et al. 1970}\hfill
\cr
2&\hbox{European sardine}\hfill & \hfill 820 & \hbox{?} & 4
& 5-8 & \hbox{C} & \hbox{Garrido et al. 2015}\hfill
\cr
3&\hbox{Black Sea turbot}\hfill & \hfill   77,000 & 74\% & 3
& 4\sim 6 & \hbox{C, {\footnotesize low accuracy}} & \hbox{Sahin 2001}\hfill
\cr
4&\hbox{Redfish}\hfill & \hfill   800 & 74\% & 10
& 12\sim 20 & \hbox{C, {\footnotesize low accuracy}} & 
\hbox{Laurel et al. 2001}\hfill \cr
5&\hbox{Siberian sturgeon}\hfill & \hfill   \sim 2,000 & \hbox{?} & 10
& 9-12 & \hbox{C} & \hbox{Gisbert et al. 2000}\hfill \cr
6&\hbox{Zebra fish}\hfill &  \hfill  12,000 & 90\% & 8
& 9-15 & \hbox{C} & \hbox{Cousin et al. 2016}\hfill
\cr
\qtb
7&\hbox{C. elegans}\hfill &  & 98\% & \hbox{\footnotesize No yolk}
& 0-1\hbox{h} & \hbox{\footnotesize Not yet done} & \hbox{}\hfill
\cr
\noalign{\hrule}
} $$
\vskip 1.5mm
\small
Notes: ``C'' means that the prediction was confirmed.
Incubation refers to the time between spawning and hatching.
The times in the columns ``End of'' and ``Predicted'' are
expressed in post-hatch days. The data given in the references
were collected for various objectives (e.g. influence of type of food,
ecotoxicology) which were quite different from the present purpose.
Cases 1-6 correspond 
to fish whereas 7 refers to a
1mm-long worm. C. elegans has a much shorter life span
than the fish: 20 days versus 15 years
for sardines (and even longer for turbots or sturgeons) which means
that days have to be replaced by by 5mn- or 10mn-long time intervals.
The scientific names are as follows:
1={\it Engraulis mordax}, 2={\it Sardina pilchardus}, 
3={\it Scophthalmus maximus}, 4={\it Sebastes mentella} and
{\it Sebastes fasciatus},
5={\it Acipenser baeri},
6={\it Danio rerio}, 7={\it Caenorhabditis elegans}.
\vskip 5mm
\hrule
\vskip 0.7mm
\hrule
\end{table}

\qI{Problems raised by the measurement of infant mortality}

Gompertz's law was discovered in 1825. 
Benjamin Gompertz was involved in the business of life insurance. 
That comes hardly as a surprise because Gompertz's law
has an obvious usefulness for various forms of life insurances. 
Ever since it was discovered in 1825 the study of Gompertz's
law has attracted considerable attention. In contrast,
the study of infant mortality was fairly neglected. 
A testimony of this neglect can be found in a paper
by Raymond Pearl and his collaborators (1941).
In the conclusion it is stated that the life curves of the
beetle {\it Tribolium confusum} ``resemble in their fundamental
pattern human life curves more closely than those of any other
organism
for which life tables have been computed''. Yet, as will
be seen shortly, the infant mortality curves of {\it Tribolium}
differ completely from human curves.
\qpar
The little interest for infant mortality curves can be attributed 
mainly to three circumstances.
\qee{1} Whereas age-specific death rates over the age of 
35 form a straight line
in a semi-log $ (x,\log y) $ plot,  infant death rates form
a straight line in a log-log plot. 
This means that the 
age-specific infant death rate is a hyperbolic power law whose
determination 
requires data points as close as possible to the
moment of birth. We will see below that for human populations
hour-by-hour postnatal death rate data are now available
but such data are relatively recent.
\qee{2} Whereas old-age mortality has not been substantially
affected by medical progress, infant mortality has been
drastically reduced over the past century. Around 1900
infant mortality during the first year of life was still
of the order of 150 per 1,000; nowadays 
in most industrialized countries
it has been reduced to around 3 per 1,000.
At first sight it might seem that a variable that is so strongly
dependent upon external factors does not have much intrinsic
biological interest. However, as will be seen below,
in spite of the huge reduction in magnitude, the shape of the
death rate has remained the same. In other words, if we write
the death rate as $ A/t^{\gamma} $, the numerator $ A $ has been
divided by a large factor, the initial point has remained
almost the same, and the exponent
$ \gamma $ has changed only slowly. 
\qee{3} Infant mortality is defined as being mortality
after birth whereas mortality occurring between conception 
and birth is called fetal mortality. Because 
for the first months after conception fetal
mortality data are very uncertain, this variable is not
considered as very significant.
Nevertheless, in ecological
studies mortality estimates usually cover the whole 
period after the production of eggs. This is for instance the
methodological option used by It\^o (1980). 
Unfortunately, with such an option survivorship
curves loose almost all significance. The reason is easy 
to understand. Many organisms, particularly insects and fishes,
produce a large number of eggs of which many die
within a short time%
\qfoot{For instance, as a fairly extreme case,
the Atlantic mackerel, ({\it Scomber scombrus}),
lays about one million eggs of which only a few survive
until 70 days after laying.}%
.
Under such conditions all 
data points for later life will be confined in a narrow range even if
one uses a logarithmic scale.\qL

\qA{Two classes of explanations}

Later we show that the decrease of
infant mortality has the same shape as in humans for various 
animal species: monkeys, lambs, birds, even crocodilians albeit
with a different exponent. However, the rule is {\it not} valid
for insects whose development proceeds through successive life stages.
For the sake of brevity, species which follow the power law decrease
will be referred to as infant mortality power law (IMPOL) species.
\qpar
For IMPOL species, this
raises the question of the origin of this
similarity. Two possible mechanisms come to mind.
\qbu It may be that IMPOL species share similar initial 
biological ``defects'' which are then ``filtered out'' through
the infant mortality process. This will be called explanation $ A $.
\qbu Alternatively, it
may be that there is a great variety of lethal effects
that may differ between individuals and species,
but that they have some properties in
common which ensure that their global effect in the course of time
will take the power law form that we observe.
This situation, which will be referred to as explanation $ B $,
would be similar to the addition of non-identical random variables 
whose global contribution, according to the central-limit theorem
of probability theory, takes the form of a Gaussian variable.

\qI{From technical devices to biological systems}

\qA{Examples of causes of failure}

When a collection of technical devices are put into operation 
at the same time $ t_0 $ a fairly high failure rate is 
usually observed during a length of time that reliability
engineers call the {\it infant mortality} phase. 
One by one, in the course of time, items which have a defect
will fail. It is important to recognize that the length of time
that it takes for defects to manifest themselves can be very variable.\qL
Let us give two examples which illustrate this point.
\qbu First, we consider incandescent light bulbs. The light
is produced by a wire filament heated to a high temperature by an
electric current $ I $ passing through it. Suppose that at some point
the section $ \sigma_1 $ of the filament is smaller than the average
section $ \overline \sigma $. 
The amount of heat produced in one second 
per unit of length of the filament is given by $ h=rI^2 $ where
$ r=\rho /\sigma $ is the the resistance per unit of length of the filament
($ \rho $ is the resistivity of  tungsten). 
A reduction in $ \sigma $ will bring about a local increase in
resistance, this in turn will produce a higher heat release 
and push up local temperature.
Around this hot spot the evaporation of tungsten will be faster.
Sooner or later, this positive feedback process 
will lead to the severance of the filament and the failure of the
lamp. \qL
As this mechanism is familiar to all physicists, one may wonder why
we explained it in some detail. The reason is very simple.
We wished to point out that there is a correspondence
between $ \sigma_1 $ and the life-time, $ \theta(\sigma_1) $, of the lamp.
Before elaborating further we wish to present a second
illustration.
\qbu Consider a device which contains one or several 
spinning wheels and suppose that one of the axes is slightly off
center. This will result in vibrations which, sooner or later,
will damage the wheel and lead to the failure of the whole device.
Here too, there will be a one-to-one correspondence between the
magnitude of the defect, $ \delta $, and the time to 
failure $ \theta(\delta) $ of the device. 

\qA{Mathematical description of the distribution of failure times}

Let us consider more closely what the existence of the
functions $ \theta(\sigma_1) $ 
or $ \theta(\delta ) $ implies
for a large sample of devices. 
In reliability studies one is usually interested in the
average life-time (also called mean time to failure or MTTF) 
of a device. In the present paper we wish to go further
in the sense that, instead of its mere average, we will study
the statistical distribution of the life-times. This means that
for a cohort of $ S_0=1,000 $ items which start to work at the same
moment, we wish to know how many will fail in the first second, 
first minute, first hour and so on.
\qpar

Mathematically, the statistical distribution of 
life-times can be described in a number of ways which are fairly
equivalent. One possible description is through the
decrease in the course of time of the number of survivors $ S=S_t $.
Although, this survivorship
description is commonly adopted especially
in biology (see for instance It\^o 1980) it is not very suggestive
because all these survivorship curves are of course decreasing
functions.
A more suggestive representation is through the evolution
of the death rate $ \mu $ as a function of age. The death rate
is defined as:
 $$ \mu(t)=\left(1/S_t\right)\left(\Delta S_t/\Delta t \right) $$
where $ \Delta t $ denotes a given age interval.
\qL
However, in medical statistics the postnatal death rate 
is defined by replacing in the previous formula
$ S_t $ by $ S_0 $:
$$ \mu_b(t)=\left(1/S_0\right)\left(\Delta S_t/\Delta t \right) $$

Usually during the infant phase $ S_t $ and $ S_0 $ are not
very different which means that $ \mu(t) \simeq \mu_b(t) $.
However when $ S_t $ becomes much smaller than $ S_0 $, the two formula
may lead to fairly different death rate shapes (see Appendix A).

\qA{Black box style explanations of the failure rate pattern}
With these definitions in mind, 
let us come back to the example of the light bulbs.
So far, we have considered one lamp, now we consider a sample
of lamps.
Suppose that all their filaments
are {\it exactly} identical with (as before) a
section that is reduced to $ \sigma_1 $ at one point.
As a result they will have the same time to failure  $ \theta_1 $
which means that for all lamps their
failure rate $ \mu_b(t) $ will be zero for $ t< \theta_1 $ 
and equal to 1 for $ t=\theta_1 $.\qL
In reality the filaments cannot be completely identical.
If the minima $ \sigma_1 $ are
distributed according to a density function 
$ f(\sigma_1) $ it will result in a density
function $ f(\theta_1) $ which in turn determines the function
$ \mu_b(t) $. So, there is a one-to-one correspondence
between the profile  $ f(\sigma_1) $ 
of the filaments and the shape of $ \mu_b(t) $.
\qpar

The data presented in the following sections suggest that
$ \mu_b(t)\sim 1/t^{\gamma} $. To explain such a fairly 
``exotic'' behavior it might be tempting to
assume a fairly complex internal structure
of the system.
For instance, in the work of Gavrilov et al. (2006)
degrees of redundancy are assumed which are
achieved by identical subsystems working in parallel.
In other papers (e.g. Peleg et al. 1998) it is assumed that there
is a whole range of failure effects, each one
characterized by a specific distribution curve.
\qpar

The light bulb example suggests that a {\it single cause} of failure
involving one parameter (e.g. the local diameter of the wire)
is sufficient to explain any shape of $ \mu_b(t) $, no matter how
exotic. Naturally, the fact that one cause of failure may be
sufficient 
does not imply that there is indeed only one cause of failure.
However, the previous argument
tells us that before turning to complex internal
structures one should rather try
to get a better knowledge of the
most likely causes of failure. 
\qpar

Conversely, it results from the previous discussion, that
the shape of the infant death rate as
a function of age gives information on the failure mechanisms 
at work 
in the system. In other words, postnatal death rates are also
an exploration tool.

\qI{Human infant mortality}

\qA{Hyperbolic power law fall of postnatal rates: 
can one predict the exponent?}

Human data are far more detailed
than for any other living organism. 

Fig. 1b and Fig. 5a,b show that the postnatal death rate 
has kept its
power law shape in spite of a considerable reduction in
the global level of infant mortality. 
This is shown for Switzerland from 1885 to 2013 and for
Britain from 1921 to 2010.
%
\count107=0  \ifnum\count107=1
\begin{figure}[htb]
\centerline{\psfig{width=17cm,figure=usaeng.eps}}
\qleg{Fig.\qhu 2 a,b\qhv Infant death rate in the death registration 
area of the United States and in England and Wales.}
{{\bf Left:} In 1910 the slope of the regression line was $ 0.65\pm 0.04 $,
whereas in 1960 it was: $ 1.01\pm 0.08 $.
The death registration area consisted of the
states which transmitted death statistics to the Bureau of
the Census in Washington. In 1910 it comprised 58\% of the US
population whereas in 1960 it comprised the whole population.
{\bf Right:} In 1921 the slope of the regression line was
$ 0.75 \pm 0.05 $ whereas in 1960 it was: $ 1.07 \pm 0.1 $.}
{Sources: USA: 1910: Mortality Statistics 1910. Bulletin 109. 
1960: Grove and Hetzel (1968, p. 210-211); Britain: Child mortality
statistics, 2013, Table 17, Office of National Statistics (UK).}
\end{figure}
\fi

%
\begin{figure}[htb]
\centerline{\psfig{width=16cm,figure=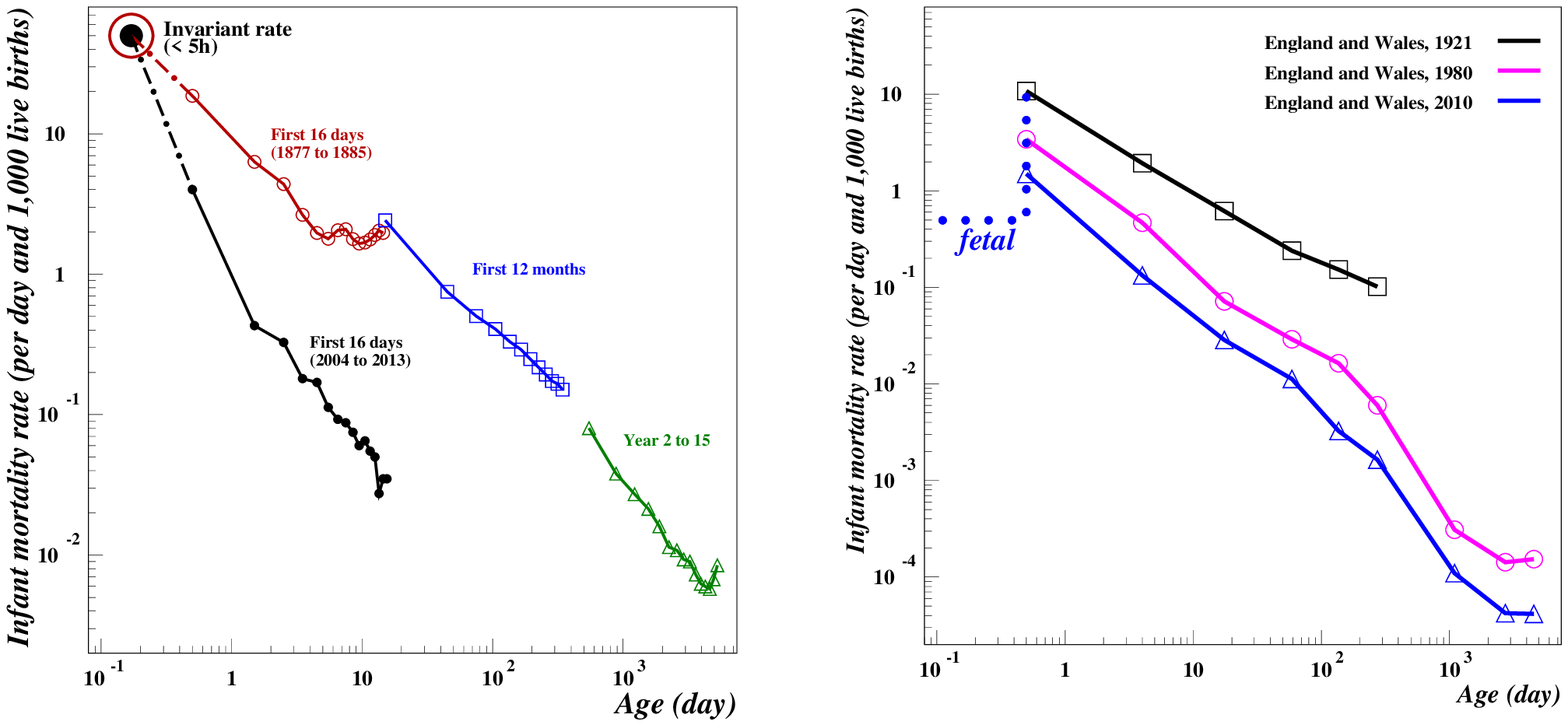}}
\qleg{Fig.\qhu 5a,b\qhv Postnatal death rates from 1 day to the 
age of 15 years.}
{{\bf Left:} Switzerland: Comparison of the two curves for the first
16 days of life suggests that the section of the curve for
times shorter than 5 hours after birth has remained unchanged
despite medical progress. The slope of the regression line
for the first 12 months is $ 
\gamma=0.85\pm 0.04 $; this slope is almost
the same as for the first 4 days; regarding the days 5 to 16,
so far we have no explanation
for why there is a level section;
for the first 15 years the slope
is $ \gamma=1.12\pm 0.14 $.
{\bf Right:} England and Wales (1921-2010). 
Comparison of the curves
shows also that in spite of a huge decrease in infant mortality,
the age-specific pattern remained basically the same.
The end of the decrease around the 
age of 10 years (i.e. some 4,000 days)
marks the limit of what in this paper we call the infant phase.
As the fetal phase would correspond to negative ages,
the magnitude of the late fetal rate is indicated in a fairly
schematic way, basically for the purpose of comparison with postnatal
rates.
From top to bottom the slopes of the regression lines are 
$ \gamma=0.75 \pm 0.05 $, $ \gamma=1.16\pm 0.12 $, 
and $ \gamma=1.19\pm 0.1 $ respectively.}
{Sources: Switzerland: The following
website of the ``Federal Office of Statistics'':\qL
{\small 
http://www.bfs.admin.ch/bfs/portal/fr/index/infothek/lexikon/lex/2.html}\qL
provides a compilation of
 historical series and in particular it contains all the annual
issues of 
``Mouvement de la population de la Suisse'' [i.e. Vital statistics
of Switzerland] starting in 1877. The data for the first 16 days
are from the volume of 1885. The data for the first 12 months are
from the ``Annuaire Statistique de la Suisse'' [Statistical Yearbook
of Switzerland] (p. 75).
Britain: Child mortality
statistics, 2013, Table 1 and Table 17, 
Office of National Statistics (UK).}
\end{figure}

The factor which determines the exponent of the power law
appears fairly clearly on Fig. 1b and Fig. 5a,b.
It is not determined by what happens immediately after
birth but rather by the death rate at the end of the 
infant mortality phase. Indeed, it can be seen that
in the close vicinity of birth (i.e. for a few hours after birth)
all death rate curves converge toward the same point.
In order to see this effect most clearly one needs series
for the same country which start as soon as possible after
birth and which are widely apart in the course of time.
The Swiss data shown in Fig. 5a turn out to best fulfill
these conditions.
\qpar

This is another instance of a fixed point
model that we
discussed in the context of old age in Richmond et al. (2016a).
Here, the exponent of the power law will become
higher when the death rate at age 10-15 decreases.
Historically, this is what happened in western countries
during the 20th century. In the perspective of
transversal analysis the same effect tells us that 
$ \gamma $ will be smaller in developing countries
than in developed countries. Testing these predictions 
may become the purpose of a subsequent paper.

\qA{How shortly after birth does the hyperbolic law start?}

Mathematically, an hyperbolic law $ 1/t^{\gamma} $ cannot hold
until $ t=0 $; there must be a cut-off time $ t_c $ prior
to which another law sets in. Fig. 6a shows that $ t_c $ is of
the order of 1 hour.

%
\begin{figure}[htb]
\centerline{\psfig{width=17cm,figure=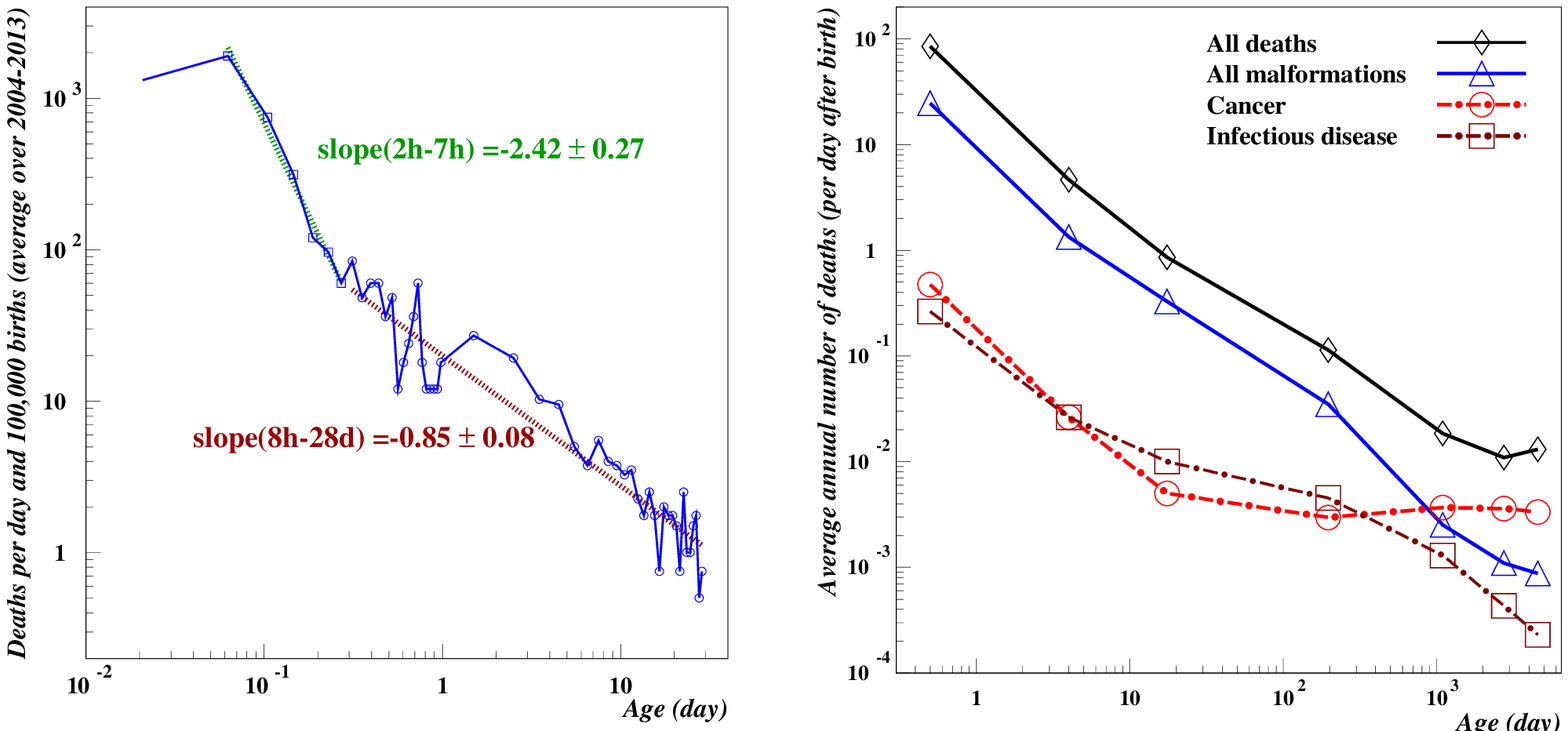}}
\qleg{Fig.\qhu 6a,b\qhv Infant death rate in Switzerland from
one hour after birth to the age of 12 years.}
{{\bf Left:} From one hour after birth to 28 days after birth
(neonatal mortality).  
If one leaves apart the first hour,
the following 7  hours are characterized by a much steeper slope than 
later times. This effect can be ``explained'' in terms of prematurity.
{\bf Right:} With its much broader
time scale this graph gives the global picture over the whole
interval marked by a decrease of the infant death rate.
The slopes of the log-log regression lines for the 3 power law
cases are as follows:
all death: $ 0.97\pm 0.12 $, malformations: $ 1.12\pm 0.07 $,
infectious disease: $ 0.69\pm 0.10 $. 
On each curve the 6 data points are averages over the following
age intervals. Days: $ (0-0.9),\ (1-6.9),\ (7-27.9),\ (28-364.9) $;
years: $ (1-4.9),\ (5-9.9),\ (10-14.9) $.
The curves refer to males and are averages over the 19 years 
1995-2013.}
{Source: Swiss Federal Office of Statistics.}
\end{figure}

We see that the end of the decrease phase occurs in the age interval
10-14 years. After that age the death rate starts to increase,
slowly at first and then after the age of 30 it assumes an exponential
growth in accordance with Gompertz's law.

\qA{Infant death rates by cause of death}

Since 1995 the
Swiss Federal Statistical Office publishes daily, weekly,
monthly and yearly infant death rates by cause of death.
The list has about 20 entries which fall into two broad
classes.
(i) Congenital malformations, 
e.g. of the nervous or circulatory system 
(ii) Diseases, e.g. infectious diseases,
cancer, diseases of the digestive system, neuropathies.
\qpar
Not surprisingly, the malformation category is largely predominant 
in the earliest part of life. As a matter of fact, the number
of deaths due to diseases is so small that 
in order to get significant estimates one needs to
add up the death numbers for all the 19 years for which
data are available.
\qpar
As an illustration let us compare deaths from cancer,
and congenital malformations. 

\begin{table}[htb]

\small

\centerline{\bf Table 3\quad Number of deaths from two causes in
Switzerland (average over 1995-2013)}

\vskip 5mm
\hrule
\vskip 0.8mm
\hrule
\vskip 2mm

$$ \matrix{
\qtb
\hbox{Cause of death}\hfill &  \hbox{Day 1} & \hbox{Year 1.0-4.9} \cr
\noalign{\hrule}
\qth
\hbox{Cancer (tumor)}\hfill & 0.42  & 5.4\cr
\qtb
\hbox{Congenital malformations}\hfill & 25  & 3.7\cr
\noalign{\hrule}
} $$
\vskip 1.5mm
\small
Notes: 
It can be seen that the death numbers move in opposite directions:
up for cancer, down for malformations. However
malformations remain important even several years
after birth.
\qL
{\it Source: Swiss Federal Office of Statistics.}
\vskip 5mm
\hrule
\vskip 0.8mm
\hrule
\end{table}

Table 3 shows that the number of deaths for the two causes move
in opposite directions; thus, clearly, they cannot be ruled
by the same law. In fact, the daily deaths by cancer first
start to decline and then level off around the age of one year.
So, why does one observe a power law for total death 
numbers? The reason is suggested by Table 3: during the first
day the deaths from malformations are 60 times more frequent
than those from cancer. Subsequently the deaths due to malformations
decrease but even years after birth, e.g.
for the age group (1 year - 4.9 years),  
they remain quite significant and indeed of the same
order of magnitude as the deaths from cancer. 
\qpar

Fig. 6b shows that unlike cancer deaths which do {\it not}
follow a power law, deaths from infectious diseases follow
a power law. It is true that there are fluctuations but there 
is no {\it systematic}
deviation. The exponent for diseases is somewhat lower than
the exponent of total deaths: 

 \hfil $ \gamma(\hbox{\normalsize infection})=0.69\pm 0.10, \quad 
\gamma(\hbox{\normalsize total})=0.97 \pm 0.12 $ \hfil 

This observation raises a question. What would be 
the shape
of the infant mortality curve during an outbreak of 
infectious disease of the kind that occurred at the end of
the 19th century? One would expect a curve that would be a
composition of malformation and infectious deaths. This
should give a power law in two parts with $ \gamma $ close
to 1 immediately after birth and then around 0.7
for ages between 2 and 10. Once data become available it will
be possible to check this prediction. 

\qA{Incidence of birthweight}

Previously we have seen that the exponent $ \gamma $ 
was not really constant but was rather declining with age.
This materialized in two ways (i) In 
Fig. 6a we saw a sharp bent after 7 hours; (ii) in Fig. 6b,
because of the averaging process, 
the decline of $ \gamma $ was smaller but
nevertheless visible.
\qpar
Fig. 6a suggests that the change in
$ \gamma $ should be attributed to prematurity. 
Indeed, in the case of pre-term birth, not only is the death rate
higher (which is hardly surprising) but also the $ \gamma $ is higher.
The reason is easy to understand. Ultimately, after a few years,
all infants, whether preterm or not, will tend to have the
same weight and same death rate. Such a convergence 
is already visible in Fig. 7a in spite of
the fact that it covers only one year.
On the other hand, the initial
death rate level of pre-term babies is in proportion of 
their degree of prematurity. Thus, inevitably, the slope $ \gamma $
must be higher for large initial death rates.
%
\begin{figure}[htb]
\centerline{\psfig{width=17cm,figure=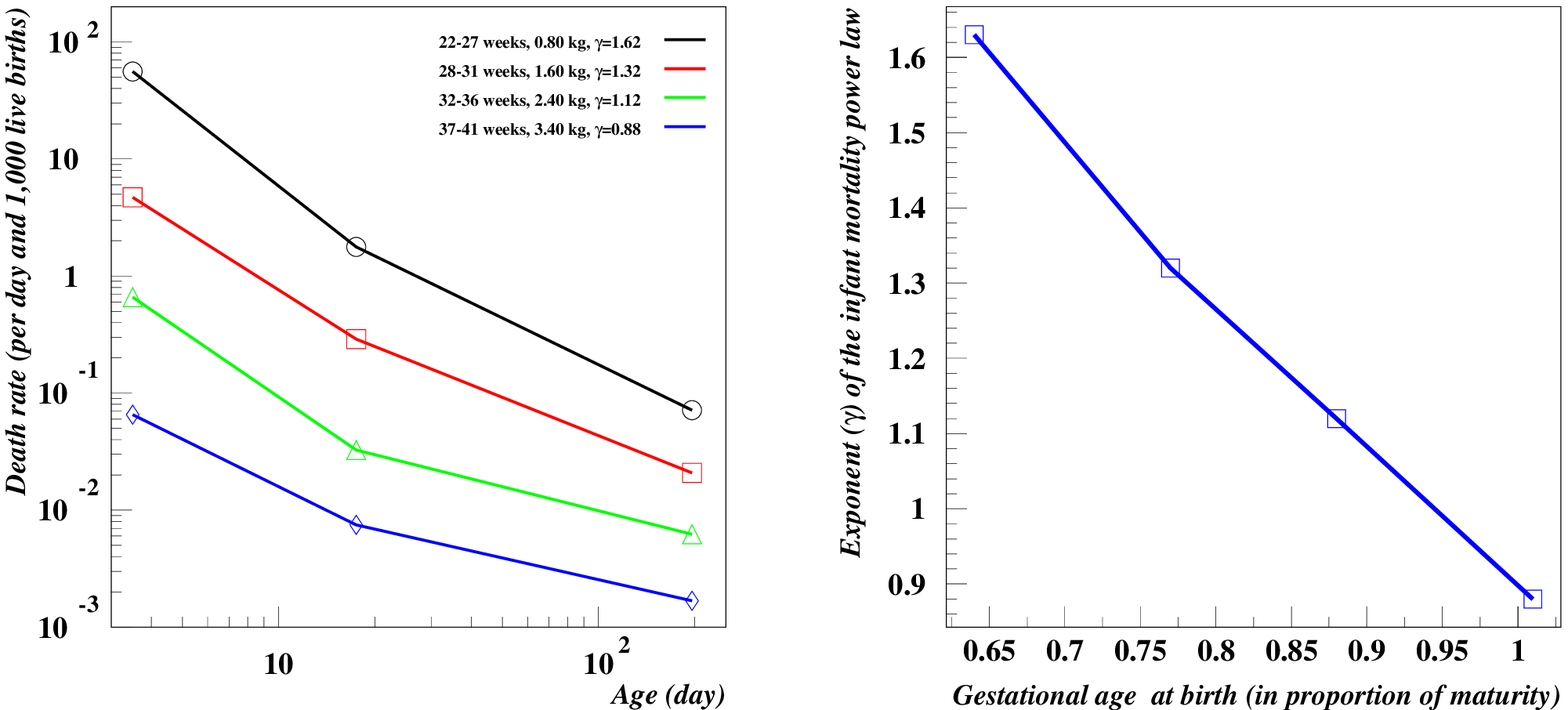}}
\qleg{Fig.\qhu 7a,b\qhv Incidence of low birthweight on death rates.}
{{\bf Left:} Low birthweight (which is usually associated with
preterm birth) results not only in higher levels of death rate
but also in higher exponents. As explained in the text, the two
effects are in fact closely connected. 
{\bf Right:} There is a linear relationship between the
prematurity index $ p $ and the exponent $ \gamma $ which takes
the form:
$ \gamma=-ap+ b,\quad a=2.0\pm 0.2,\ b=2.90\pm 0.02 $.
}
{Source: Swiss Federal Office of Statistics.}
\end{figure}

Fig. 7b  summarizes the relationship between prematurity and
the exponent $ \gamma $. 
The prematurity index
used for the horizontal axis of Fig. 7b was defined as:
$ p=\hbox{(gestational age at birth)/(9 months)} $. Then,
the regression line follows the equation:
$$ \gamma=-ap+ b,\quad a=2.0\pm 0.2,\ b=2.90\pm 0.02 \qn{1} $$

This relationship is defined with high accuracy as shown
by the small size of the error bars for $ a $ and $ b $.
\qpar

It would be really surprising that a relationship that holds
with such accuracy for humans
would not hold as well for other mammals. 
Of course, the coefficients
$ a $ and $ b $ will not be exactly the same but one would
expect a linear relationship to hold with good precision.
\qpar

The next step would be to
set up an experiment specially designed to 
observe this effect and to measure $ a,b $ with good accuracy. 
Until this is done, we must rely on makeshift data extracted
from experiments which were conducted for a completely
different purpose. As an illustration, one can
mention an experiment described in a paper by Price et al. (1972).
In this paper the authors measured the death rate of monkeys
({\it Macaca mulatta, Macaca fascicularis, Macaca artoides})
at age $ 0-8 $ days and $ 8-30 $ days for different classes
of birthweights. As the number of births is not very large
(only 91 for the 3 species) the data shown in the paper
display large random fluctuations. 
By lumping together successive
birthweight classes, keeping only two, the
fluctuations could be reduced substantially. As a result, one gets
a relationship similar to (1) with the following coefficients:

\hfil $ \gamma=-ap+ b,\quad a=4.3,\ b=5.8  $ \hfil
 
There are probably substantial error bars for $ a,b $ but
they could not be estimated because the regression line
is defined by two points only.
\qpar

The following sections are devoted to non-human living organisms.
We will proceed from the cases that are closest to humans
such as farm mammals and primates to more remote cases 
such as plants and trees.

\qI{Farm mammals: piglets and lambs}

\begin{figure}[htb]
\centerline{\psfig{width=6cm,figure=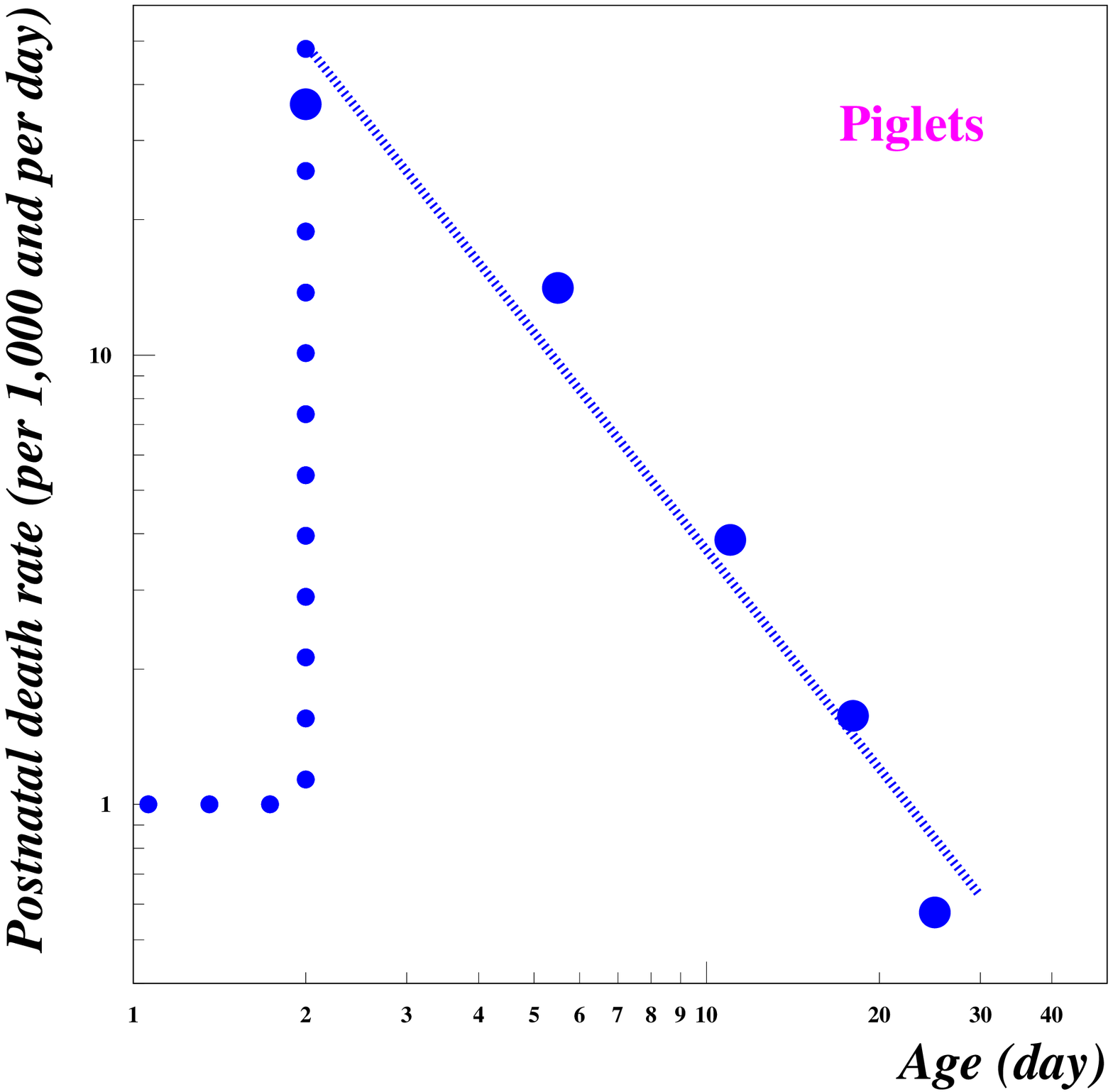}}
\qleg{Fig.\qhu 8\qhv Postnatal mortality of piglets.}
{The slope of the regression line is $ -1.6\pm 0.4 $.
The data are from a study of the US ``National Animal Health
Monitoring System'' (NAHMS) for 1990. The dotted
part of the curve is a schematic representation of the
spike (age 0 and negative
ages can of course not be displayed on a log-scale).
As fetal death rates
of piglets 
are not well documented the level was set to a value
similar to late fetal rate in humans.}
{Source: US Dept of Agriculture 1992.}
\end{figure}

For farmers early deaths of farm animals represent a 
substantial economic loss. In the United States and Australia
about 16\% of the new born piglets died before
reaching the age of one month. This led to studies done
by the Departments of Agriculture whose main aim 
was to understand the causes of death%
\qfoot{For instance the study about lamb mortality
revealed that overcrowded pens lead to a high mortality due
to the ewes laying on their lambs. This cause of death
represented 25\% of the deaths which occurred from
day 1 to day 7.}%
.

\begin{table}[htb]

\small

\centerline{\bf Table 4\quad Neonatal mortality of piglets and
lambs}

\vskip 5mm
\hrule
\vskip 0.7mm
\hrule
\vskip 2mm

$$ \matrix{
\hbox{Case}\hfill &  \hbox{Number of} & \hbox{Exponent of}&
\hbox{Correlation} \cr
\qtb
\hbox{}\hfill &  \hbox{age intervals} & \hbox{power law}&
\hbox{(log-log)} \cr
\noalign{\hrule}
\qth
\hbox{Piglets}\hfill &   & &\cr
\hbox{\quad USA, 1992}\hfill & 5  & 1.60\pm 0.40 & 0.98\cr
\hbox{\quad Australia, 1976, including still-births}\hfill & 8  &
1.20\pm 0.25& 0.97\cr
\hbox{\quad Australia, 1976, excluding still-births}\hfill & 8  &
0.98\pm 0.34& 0.92\cr
\hbox{Lambs}\hfill &   & &\cr
\qtb
\hbox{\quad USA, 1997}\hfill & 10  & 1.00\pm 0.30& 0.92\cr
\noalign{\hrule}
} $$
\vskip 1.5mm
\small
Notes: Neonatal and pre-weaning correspond approximately
to the same period of time after birth, namely about 30 days.
All three surveys are
large scale studies involving several thousands births.\qL
{\it Sources: Piglets: US Dept of Agriculture 1992, Glastonbury 1976.
Lambs: Berger 1997.}
\vskip 5mm
\hrule
\vskip 0.7mm
\hrule
\end{table}

As a by-product these studies gave death rates by age
which allowed us to check whether they follow a power law or
not. The power law shape was confirmed (Fig. 8)
with exponents summarized in Table 4.
\qpar

Apart from the studies concerning farm animals there 
are few sources from which one can get accurate data
about infant mortality for animals. One other important
source consists in the records from zoological gardens.
It is this source that will be used in the two following
subsections.

\qI{Primates}

Primates are one of the main attractions of zoos
and it is not surprising therefore that major zoological gardens
have large populations of species of this taxon%
\qfoot{A taxon is a group of species with some common
characteristics. More details can be found in Campbell
and Reece 2004, p. 767-769.}%
. 
Primates have the additional interest
of being a kind of stepping-stone between humans
and other, more distant, mammals.  For all these reasons
primates warrant a close look. 

\qA{Postnatal death rates for various species of primates}

As the detailed data that we needed
were not available%
\qfoot{Later on we will use some data published in Kohler et al.
(2006); these data have the advantage of being available for a
broad range of species but they have a poor time resolution
in the sense that they give infant mortality data only for 
the ages of one week and one year.}
one of the authors (V.P.) conducted
a special investigation using registers from the London Zoo
called {\it Daily Occurrences}. These records are available at the
Archives of the Zoological Society of London. These data record 
the arrivals and departures of animals on a daily basis. It can be added
that in a general way primates get particularly close attention. 
\qpar
Fig. 9a shows that all small primates investigated follow
similar power laws with an exponent which, on average is equal
to $ \gamma=1.24\pm 0.2 $.
%
\begin{figure}[htb]
\centerline{\psfig{width=14cm,figure=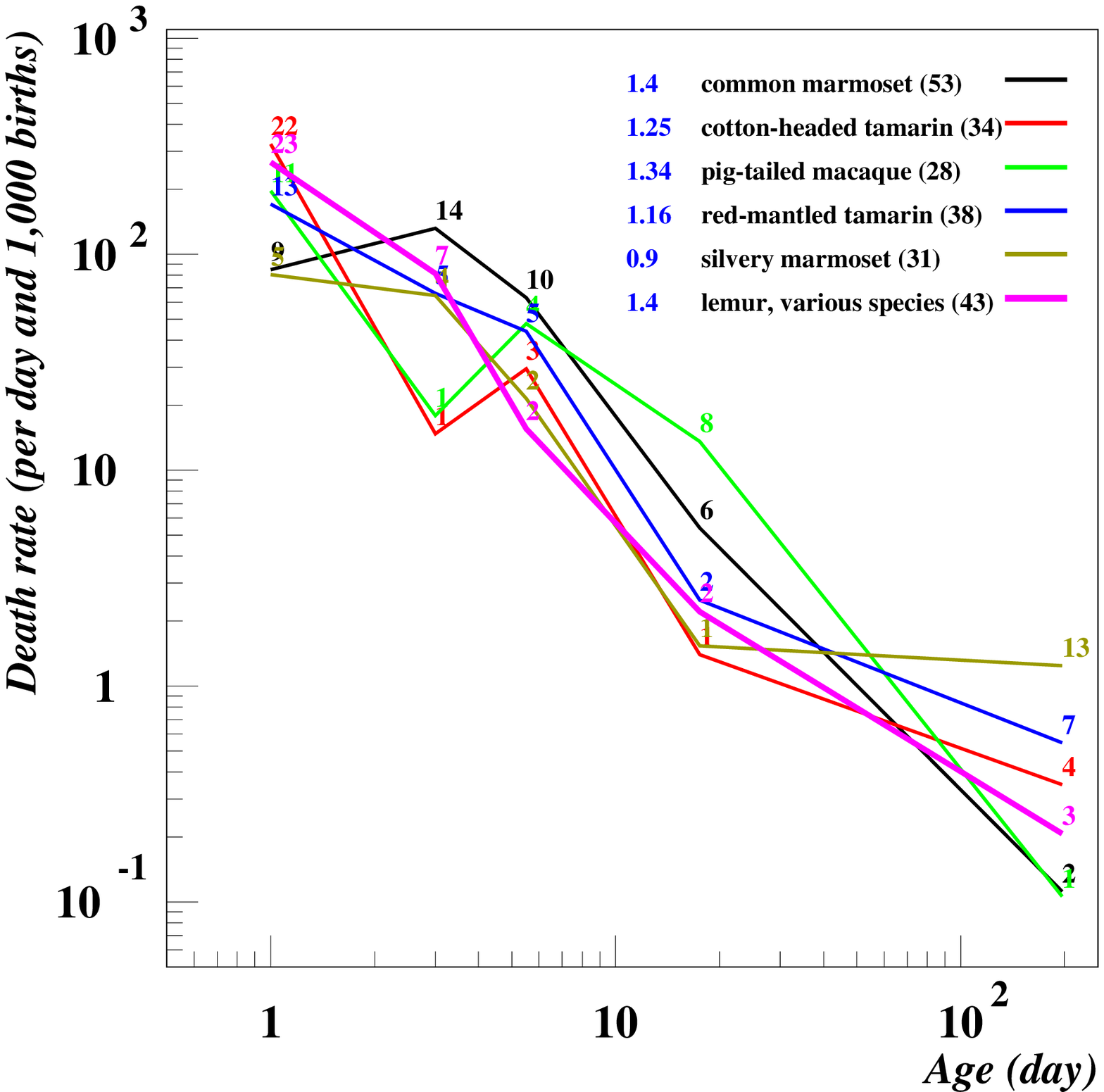}}
\qleg{Fig.\qhu 9a\qhv Infant mortality rates of primates.}
{The numbers which follow the names of the
species give the size of the subgroup of individuals whose
birth and death dates were recorded in the {\it Daily
Occurrences} volumes of the London Zoo in the period 1970-2000.
The numbers printed on the curves give the deaths in each
age interval. Finally the numbers (in blue) which precede
the species names are the exponents $ \gamma $ of the 
power law.}
{Source: Archives of the Zoological Society of London.
For more information about these data, please contact
Ms. Violette Pouillard.}
\end{figure}
%
\begin{figure}[htb]
\centerline{\psfig{width=8cm,figure=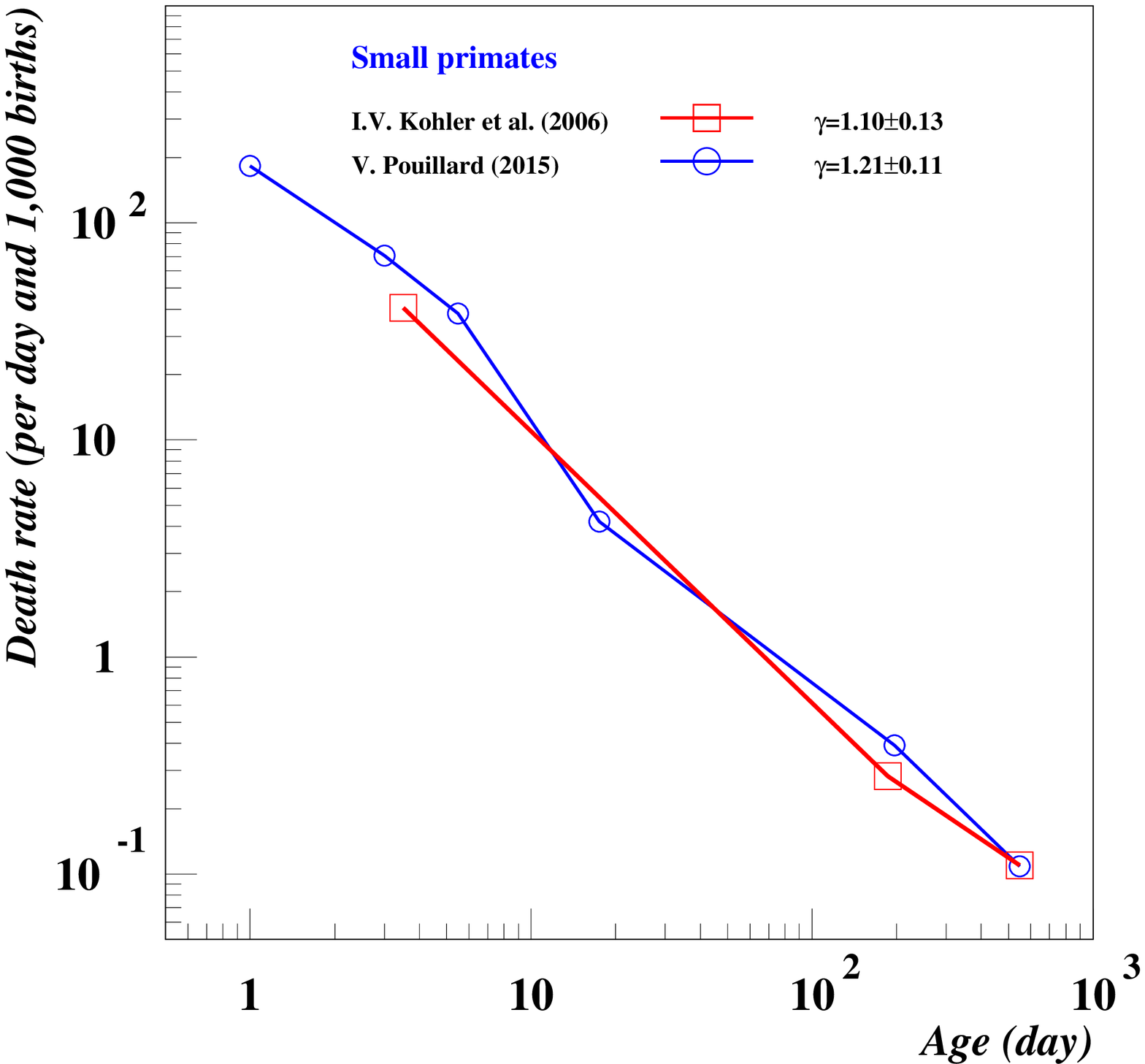}}
\qleg{Fig.\qhu 9b\qhv Infant mortality rates of primates.}
{For the whole group of small primates
the graph provides a comparison between two separate
data sets. It is reassuring to see that they show a good agreement.
The comparison provides a welcome confirmation
of the data of Kohler et al. (2006), a check that
is all the more useful
due to the fact that they consist in only 3 data points 
of which only 2 can be considered as pertaining to infant mortality.}
{Sources: Archives of the Zoological Society of London,
Kohler et al. (2006).}
\end{figure}

\qA{Consistency of separate data sets}

In physics it is the rule that an experiment done by one team
is repeated%
\qfoot{Yet, never {\it exactly} in the same way in the sense
that the devices and measurement methods
are not the same. Taken together, these experiments
will show what are the necessary and sufficient conditions
for observing the effect under consideration. In other words,
these semi-repetitions define the envelop of parameters
within which the effect occurs.}
and checked by one or several other researchers.
In biology this is fairly rare and in the social sciences it
is exceedingly rare. Here, however, we have the opportunity
to carry out such a comparison for the case of small primates. 
The two observations were performed
independently and rely on data that differ
in several respects.
\qbu The time periods are not the same: 1970-2000 versus 1998-2003.
This has an incidence on the manner of recording the data.
During the 1970-2000 period the data were mostly
recorded by hand whereas 1998-2003 already
belongs to the digital era. 
\qbu The zoos are not the same: London zoo versus various zoos
in North America and Europe. Not only are the animals not the
same but even the species are not exactly the same although there
is of course a broad overlap. The species belonging to
the ``small primate'' category considered by
Kohler et al. are listed on p. 416 of their paper.
\qbu Finally, the sources of the data are not the same:
{\it Daily Occurrences} of the London zoo versus ``International Species
Information System'' (ISIS). We will say more about ISIS in
a moment. 
\qpar

In short, the quasi-coincidence of the two data sets
is a good testimony of the robustness of this kind of observations.
In particular it shows that despite limited data points
the observations collected compiled by 
Kohler et al. (2006) reflect fairly well
the overall shape of the infant mortality rates as a function of
age.
\qpar

A second confirmation can be mentioned. 
In a paper by Hird et al. (1975)
we are told that ``the neonatal mortality rate 
 was 10.8\%, and the post-neonatal mortality
rate (deaths between 31-183 days) was 6.9\%. These data allows us to
compute $ \gamma $. One gets: $ \gamma=1.06 $ which is quite
consistent with the previous results.
\qpar

Finally, it must be mentioned that
we also came across a paper whose results are 
inconsistent with the previous ones. The study by Shaughnessy
et al. (1978) gives the following distribution of deaths
over the 4 first weeks after birth: $ w_1=24\%,\ w_2=59\%,\
w_3= 17\%,\ w_4=9.6\% $. The fact that $ w_1<w_2 $ is in
contradiction with all observations that we have reported so far.
Secondly, although
$ w_2,w_3,w_4 $ follow a power law, the corresponding
exponent is equal to $ 2.6 $ which is much higher than
the exponents seen so far. These data relied on births
which occurred between 1966 and 1972 at the breeding
colony of the ``Litton Bionetics'' company in Maryland.
One is tempted to think that this population of 
{\it Macaca mulata} was ``special'' in some respect.
In order to get a clearer understanding,
additional information would be needed about the conditions
prevailing at that breeding center.

\qI{Broader view of animals kept in zoos}

The dataset used in the paper by Kohler et al. (2006) was
not designed to study infant mortality.
This is shown clearly by the fact
already mentioned that there are only 3 (very distant)
data points: 1 week, 1 year and 2 years. 
Without the control test performed in the previous subsection it
would have been hazardous to use them to study
postnatal mortality. However,
encouraged by the consistency seen for small primates
we have considered other subgroups. The results are summarized
in Fig. 10.
%
\begin{figure}[htb]
\centerline{\psfig{width=13cm,figure=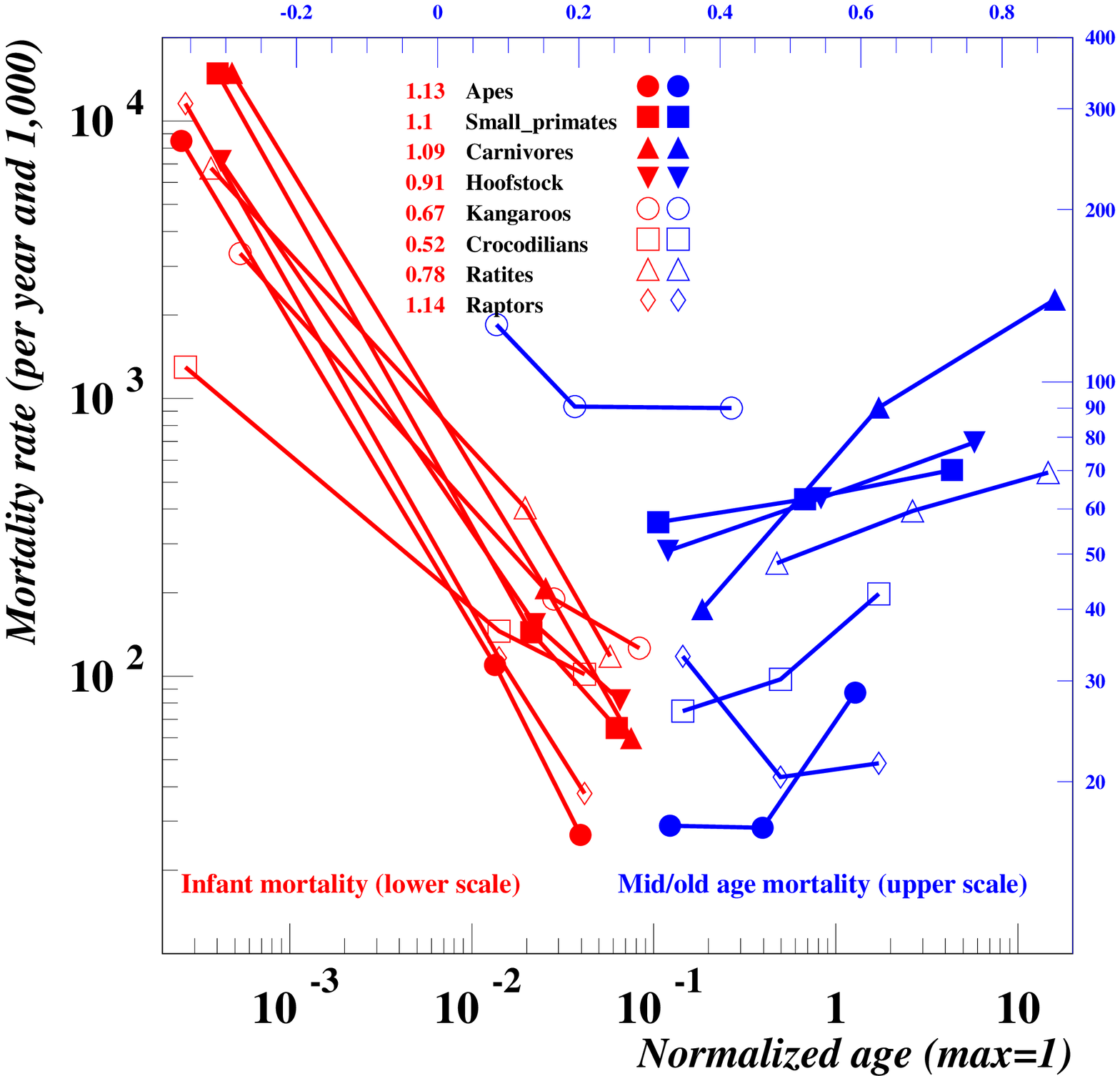}}
\qleg{Fig.\qhu 10\qhv Infant mortality rates versus mid-age rates.}
{The scales for the mid/old age rates are given by the axis on the
right-hand side and the top axis.  
The numbers which precede the names of the subgroups are 
the exponents $ \gamma $ of the power law for the infant death rates.
Despite the fact that
there are only 3 data points, thanks to the good correlations,
the error bars (at probability level 0.95) are only about
$ \pm 8\% $ on average.}
{Source of the data: Kohler et al. (2006).}
\end{figure}

One can learn two things from this graph.
\qee{1} The infant death rates are much more regular and uniform
than the death rates of mid/old age (we come back to this point
below). They are also much higher.
\qee{2} With respect to the exponent $ \gamma $,
of the 8 subgroups there are two which emerge as quite different,
namely kangaroos and crocodilians. The average
of the 6 others is (with probability level 0.95):
$ \gamma_m=1.02\pm 0.12 $. The case of the crocodilians is
particularly spectacular because they have not only
a lower slope but also a much lower overall death rate.
\qpar

Why do the mid/old age rates show such a high dispersion?
Two answers come to mind.\qL
It may be due to the well known fact that aging
processes are very diverse, as shown
by the classification of
survivorship curves into three types I, II, III,
not to mention all kinds of intermediate types.\qL
There may be a second reason. A look at 
the ISIS records shows that most of the animals do not
spend their whole life in the same zoo; instead they are
repeatedly loaned (or sold) by one zoo to another.
This may be disturbing for the animals but it raises
also a book keeping problem. In the paper
by Kohler et al. (p. 429) it is reported that even
highly visible animals such as gorillas ``are assigned multiple
identification numbers in various regions around the world''.
What occurs for gorillas is also likely to occur for 
other species. The only difference may be that for low
profile species the inaccuracies in the records remain
unnoticed.

\qI{Insects}

Before becoming adult, insects
go through several life stages: (i) embryo, (ii) larva, 
(iii) pupa (or nymph which has already the form of an adult
but not yet its size)
and (iv) finally adult. According to
our definition of the infant mortality we will
consider only the adult stage, i.e. the stage
which leads to sexual maturity. The nymphal stage
should be included in the adult stage because 
the nymph has already all the properties of the adult.
\qpar

In each such transformations the insects
are exposed to changing external situations. From
caterpillar (larval form)
to butterfly there is clearly a drastic 
change in environmental conditions. In other words,
there are several filtration processes
which may
be independent or inter-dependent. 
\qpar
Now, let us have a look at some data.

\qA{Successive instars}

The best test is provided by successive instars.
It can be recalled that in its nymphal stage, an insect has 
already its adult shape but because of the rigidity of its
outside cuticle, in order to achieve
its adult size, it must undergo several molts. The phase between
two successive molts is called an instar. Often there are up
to 5 or 6 instars. The last instar is the adult. It\^o (1980)
gives several survivorship curves on which successive
instars are plotted. It appears that often the screening between
the first and second instar is weaker (in the sense of a smaller
$ \mu (t) $) than the screening following the
hatching which produced the first instar. As an illustration
one can mention the case of {\it Mogannia minuta} (sugar cane
cicada) as particularly clear (p. 84). As the survivorship
curves drawn by It\^o were not designed to study this specific
issue, they do not have the required accuracy in the sense
that most instar
death rates are compressed within a narrow interval. One would need
the original data. We leave this question open for further
examination.

\qA{Beetle}

Finally, we consider the case of a beetle, namely
``{\it Tribolium confusum} Duval'' (Pearl et al. 1941).
%
\begin{figure}[htb]
\centerline{\psfig{width=8cm,figure=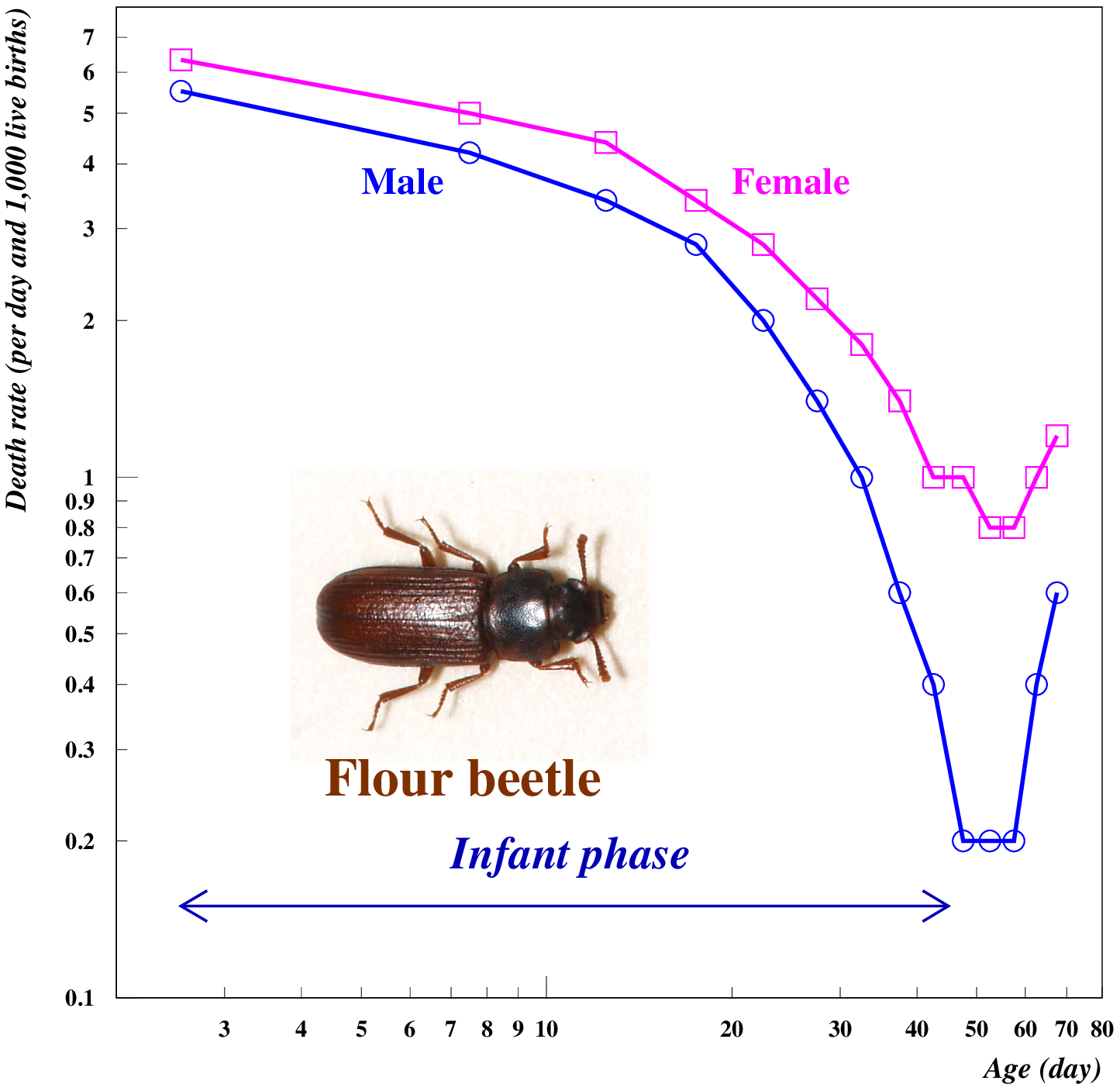}}
\qleg{Fig.\qhu 11\qhv Infant death rate of a flour beetle
({\it Tribolium confusum}.}
{In the present case, unlike previous cases,
$ \mu_b $ does not have a steep downward
slope after ``birth''. Birth, here, in fact means emergence
from the pupal stage. This stage follows a previous
life stage as a larva in the form of a worm.
Thus, there are several successive filtration processes.}
{Source: Pearl et al. (1941, p. 13-14)}
\end{figure}
%
It can be noted that (as shown in Pearl's paper)
the survival curves of males and females in late age
are very different. The 50\% proportion is reached
for 170 versus 210 days respectively whereas the 10\% proportion
is reached for 410 versus 375 days respectively.
In contrast during their infant phase, male and female rates
display parallel changes. This, once again, illustrates
the fact that infant mortality is ``simpler'' (in the sense
of being less affected by exogenous factors) than aging.\qL
The fact that Fig. 11 does not display a power law suggests
that a different mechanism is at work.

\qI{Plants and trees}

In its principle, the process which leads from a tiny embryo to
a seedling does not much differ from what we see in 
the growth of fish or mammal embryos. 
It is marked by similar steps
of division and differentiation. For many (yet not all) plants
the growth of the embryo continues for a given time interval after
the formation of the seed. This means that germination cannot
occur immediately. If, for some reason, this delay
is not respected the seedling will face a prematurity problem
just as in human premature births. 

\qA{Yolk sac effect in plants}

In the time interval between germination and
formation of roots and leaves, the nutrients contained in the
seed play the same role as the yolk sac for fish larvae%
\qfoot{For instance, according to Hanley et al. (2004),
this phase lasts about 11 days for
sunflowers ({\it Helianthus annuus}) and 12 days for pea
({\it Pisum sativum}). Thus, one should observe
a mortality spike in the interval 10d-15d 
after germination. 
Its amplitude may be small (perhaps about 1\%) but
should be visible on samples of 1,000 seeds or more.}%
.
This parallelism is described in the following lexikon.

\begin{table}[htb]

\small

\centerline{\bf  \quad Lexikon of development terms for
fish versus plants}

\vskip 5mm
\hrule
\vskip 0.7mm
\hrule
\vskip 2mm

$$ \matrix{
\qtb
\hbox{\bf \color{blue} Fish:} \hfill&\hbox{embryo} \hfill &\hbox{egg}\hfill
& \hbox{hatching}\hfill&\hbox{yolk sac}\hfill 
& \hbox{larva} \hfill & \hbox{young adult}\hfill \cr
\noalign{\hrule}
\qth
\hbox{\bf \color{blue} Plant:} \hfill&\hbox{embryo} \hfill &\hbox{seed}\hfill
& \hbox{germination}\hfill&\hbox{endosperm}\hfill 
& \hbox{seedling} \hfill & \hbox{juvenile}\hfill \cr
\qtb
\hbox{} \hfill&\hbox{} \hfill & & &\hbox{cotyledon}\hfill 
& \hbox{} \hfill & \hbox{}\hfill \cr
\noalign{\hrule}
} $$
\vskip 1.5mm
\small
Notes: Despite the different vocabulary there is a strong
parallelism between the successive phases.
The two terms given for the yolk sac phase reflect the
difference between plants whose seed germinate under or over the
surface of the soil (they are called hypogeal or epigeal species).
\vskip 5mm
\hrule
\vskip 0.7mm
\hrule
\end{table}
Therefore one would not expect the death spike to occur 
right at germination but rather when the nutrients contained
in the seeds are exhausted. The accuracy of the data that
we were able to find so far is too low to allow this
prediction to be tested.

\qpar

\qA{The hyperbolic power law effect}

Plants are simpler than insects whose development
goes through several
successive stages. Moreover, field
observations of plants are easier to perform
and are more reliable
than observations on animals for the obvious reason that unlike
animals, plants do not move around%
\qfoot{However, for plants
a major difficulty is how to define the moment
of death. For plants which exhibit positive phototropism.
the termination of effect may be used to define death in a 
more accurate way than just by its aspect.}%
. 
Yet, surprisingly, very few life tables have
been set up for plants. In 1975, Valen wrote ``Although
the study of survivorship [curves] has been an important part
of animal ecology for 30 or 40 years, there are few studies
on plants''.  The same observation still basically
holds in 2015. 

\qA{Field observation of palm trees}

The results 
for the palm trees {\it Enterpe globosa} studied by Valen are
summarized in the graph of Fig. 12. 
Two features are of particular interest.
\qbu As for humans, the
age interval during which the death rate decreases
roughly coincides with the period before sexual maturity.
\qbu The death rate follows a power law fairly accurately.
The $ (\log t,\log \mu_b) $ correlation is $ 0.998 $. 
Yet, with an exponent
equal to $ \gamma = 2.6\pm 0.4 $ the decrease is
almost three times steeper than in previous cases. 
Is this property shared by other big trees? 
A further discussion can be found in Appendix C of the
version of the paper available on arXiv.
%
\begin{figure}[htb]
\centerline{\psfig{width=8cm,figure=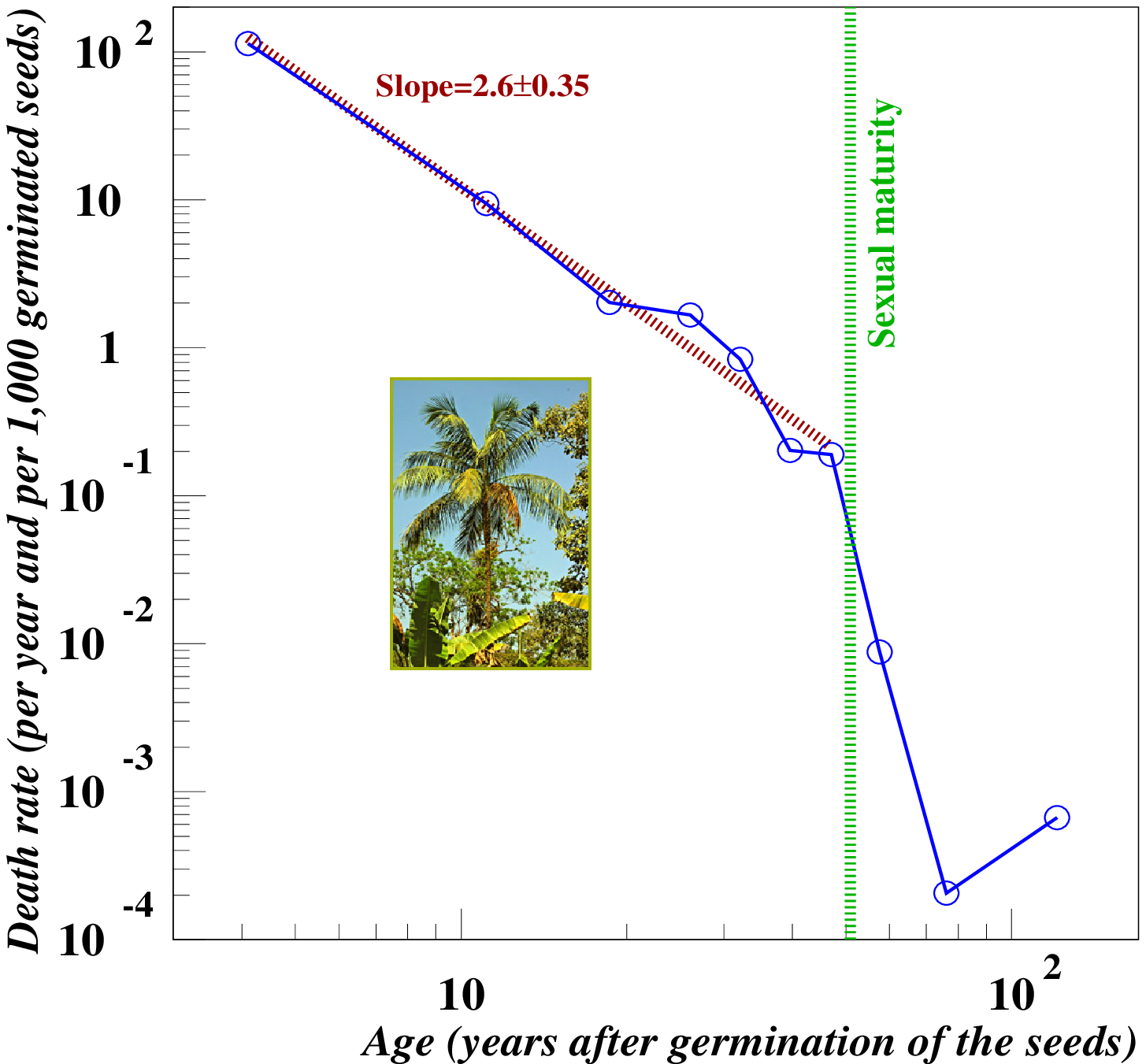}}
\qleg{Fig.\qhu 12\qhv Infant death rate of a palm tree.}
{This palm tree ({\it Euterpe globosa}) lives about 200
years and reaches a height of 20 meters. The observations
were made in Puerto Rico. Instead of being close to one as for most
other cases, it is close to 2.5. At this point we do not
know why.}
{Source: Valen (1975, p. 263)}
\end{figure}

\qI{Conclusion}

\qA{Main results}

What we learned in this paper can be summarized in the following
observations. 
\qee{1} {\bf General}\quad 
(a) Infant mortality (in the sense of being a phase
during which the death rate {\it decreases})
is an ubiquitous phenomenon in
living organisms. It was shown to 
exist in mammals, fish and plants.
(b) In all cases for which appropriate data could be found,
we have seen that there is
a hyperbolic power law 
(i.e. $ 1/t^{\gamma} $ starting in the vicinity of $ t=0 $
and then holding for large $ t $)
decrease of the death rate. 
(c) The exponent $ \gamma $ is usually close to one.
The plant instances in which we found exponents
as high as 3 need to be confirmed by
observations under
controlled conditions (i.e. no predating, appropriate
supply of water, and so on).\qL
(d) For most animal species it appears that
there is much more regularity in infant death patterns
than in old-age death patterns.
\qee{2} {\bf Yolk sac larvae}\quad 
The {\it Transient Shock} conjecture
offered predictions regarding the existence and
timing of death rate spikes and these
predictions were confirmed by observation. 
\qee{3} {\bf Humans}\quad (a) What we called the hyperbolic
power law 
starts to hold in the second hour following birth. 
(b) The fact that in the hours following
birth $ \gamma $ is of the order of 2 rather than 1 can be explained
by the effect of premature birth.
(c) There is a negative correlation 
between $ \gamma $ and the mortality rate in the
10-14 age group.
(d) If one makes a distinction between different
causes of death, most of them
follow a power law, yet not necessarily with the same exponent.
For instance, death due to infectious disease is characterized
by an exponent of 0.69 instead of 0.97 for the all-causes curve.
(e) There is a linear relationship between the index
of prematurity and the exponent $ \gamma $.
\qee{4} {\bf Primates}\quad For primates (and more generally
for any mammals for which data are available) one observes
the same infant mortality pattern than for humans.
\qee{5} {\bf Multi-stage organisms}\quad 
For insects in their adult life stage
the mortality rate does also decrease with age
but it does not follow a power law. This may be due to the fact
that this stage was preceded by several
selection processes (larva, pupa, nymph).

\qA{Origin of the hyperbolic power law of mortality decrease}

There is one question for which we have no answer so far, namely
how can one explain the origin of the power law decrease 
observed for the
infant mortality rate, a feature that is common to so many species.
Two types of explanation come to mind.
\qbu First, there is the filtering effect of defective individuals
that we have already discussed (explanation of type A).
This can be referred to as a
static explanation in the sense that it does not assume any
transformation taking place in individual organisms.
A tentative model is outlined in the arXiv version of the
present paper.
\qbu Then, there is a dynamic explanation in which 
one assumes an improvement of the immune system in the course
of time.
It is the immune system which ensures survival in later life;
in addition we
know that vaccination actually works;
therefore it seems natural to assume that the effectiveness of the
immune system improves as individuals interact 
with the outside world and
experience in small
doses various ``nasty'' microorganisms which trigger 
the emergence of
antibodies. We also know that this
process takes time for at birth the immune system is largely
under-developed and newborn individuals have to rely on what is 
called passive immunity, namely 
maternal antibodies transmitted to them.
\qpar

In order to substantiate the second explanation one would
need age-specific data for vaccination effectiveness%
\qfoot{It is defined as $ E=1-I_v/I_{nv} $ where $ I_v $ is the
incidence of the disease under consideration (e.g. influenza)
in the vaccinated population and $ I_{nv} $ is the incidence
in the population that is not vaccinated. If not a single
vaccinated person gets the disease, $ E $ will be equal to 1.}%
.
An alternative metric would be to measure
the antibody concentration in the blood of animals
(which were not subject to vaccination) 
as a function of their age.

\qA{Agenda for future research}

We wish to stress the fact 
that for various organisms such as plants, bacteria,
microorganisms, insects
it would be relatively easy to set up infant mortality experiments
because, in contrast with the study of aging, the observation
time can be much shorter.
Our plan is to build a chain of systems
which starts from the simplest (e.g. technical devices, plants,
primitive animals such as {\it C. elegans}) 
and progressively embraces more complicated
systems. Such a program was already considered at the end
of the 19th century by Alfred Espinas (1878). As a matter
of fact, at that time it seemed to be a fairly natural idea to
complement sociological investigations with studies about
other living organisms. Nowadays such an approach has become
fairly uncommon.
\qpar
As an example of their interest, such
experiments may shed new light on
the hierarchical complexity structure of living
organisms. 
Here again a parallel with technical systems
may help to explain this idea.
\qpar

A modern airliner is made up
of many functional components (wings, engine, computer and so
on). These components comprise large numbers of
smaller elements (screws, electronic chips and so on).
In the last step of the building process
the functional components are put together on the assembly line.
Finally, tests are performed with the
purpose of detecting possible defects. \qL
In principle deficiencies may occur at the three levels:
small elements, components, assembly line. 
However, observation shows that the small elements 
have a very low defect rate. This is fairly understandable
because they are produced through standard
manufacturing processes and are fairly easy to 
control.
Similarly cell division seems to be a very reliable generation
process. There are some $ N=3\times 10^{13} $ cells in the human
body (Bianconi et al. 2013, note that a large proportion of 
them are red blood cells). On average 1\% of them must be replaced
every day. This raises the question of what defect rate
is acceptable. The answer is certainly highly organ dependent.
A defect rate of 1 per 1,000 may be quite acceptable for red blood
cells but may not allow an organ such as an eye to work
properly. Needless to say, 
a component may be defective even though
all its elements are good. Similarly, if mistakes
are made at the assembly line level the aircraft
may be defective even though all its components are 
flawless.
\qpar
A careful analysis of postnatal death rates 
across various species may give
useful information about this multi-level organization

\appendix

\qI{Appendix A: Infant mortality data}

In this Appendix we first discuss the definitions of 
mortality rates. Then, we explain the conditions under
which infant mortality should be measured.
Finally, we give some indications about possible 
data sources.

\qA{Definitions}

In statistical sources death rates are computed in two different
ways depending on whether they concern infant mortality 
(defined in the broad way of post-natal mortality 
used in this paper) or not.
The definitions are recalled in Fig. A1a. 
This can create a good deal of confusion because a given
survival function will lead to different mortality patterns
depending on which definition one uses.
This is illustrated in Fig. 1b for
an exponential and a power law fall.

\begin{figure}[htb]
\setbox48=\vbox{\hsize=16cm
{\large
\centerline{Definition of standard versus postnatal death rates}
$$ \matrix{
t:\hbox{ \large age } (u=\log t) \hfill & &\cr
s(t):\hbox{ \large survivors at age } t \hfill  & &\cr
\mu(t):\hbox{ \large standard death rate} \hfill & & \mu=|\left( 1/s
\right)\left( ds/dt
\right)| \hfill & y=\log\mu \hfill \cr
\mu_b(t):\hbox{\large postnatal death rate} \hfill & & 
\mu_b=|(1/s_0)(ds/dt)| \hfill &  y_b=\log\mu_b \hfill \cr
} $$
}
}
$$  \boxit{8mm}{\box48} $$
\qleg{Fig.\qhu A1a\qhv Standard death rate ($ \mu $) and
postnatal death rate ($ \mu_b $).} 
{The logarithms of the death rates,  $ y, y_b $,
are used below in log-log plots.}
{}
\end{figure}

%
\begin{figure}[htb]
\centerline{\psfig{width=16cm,figure=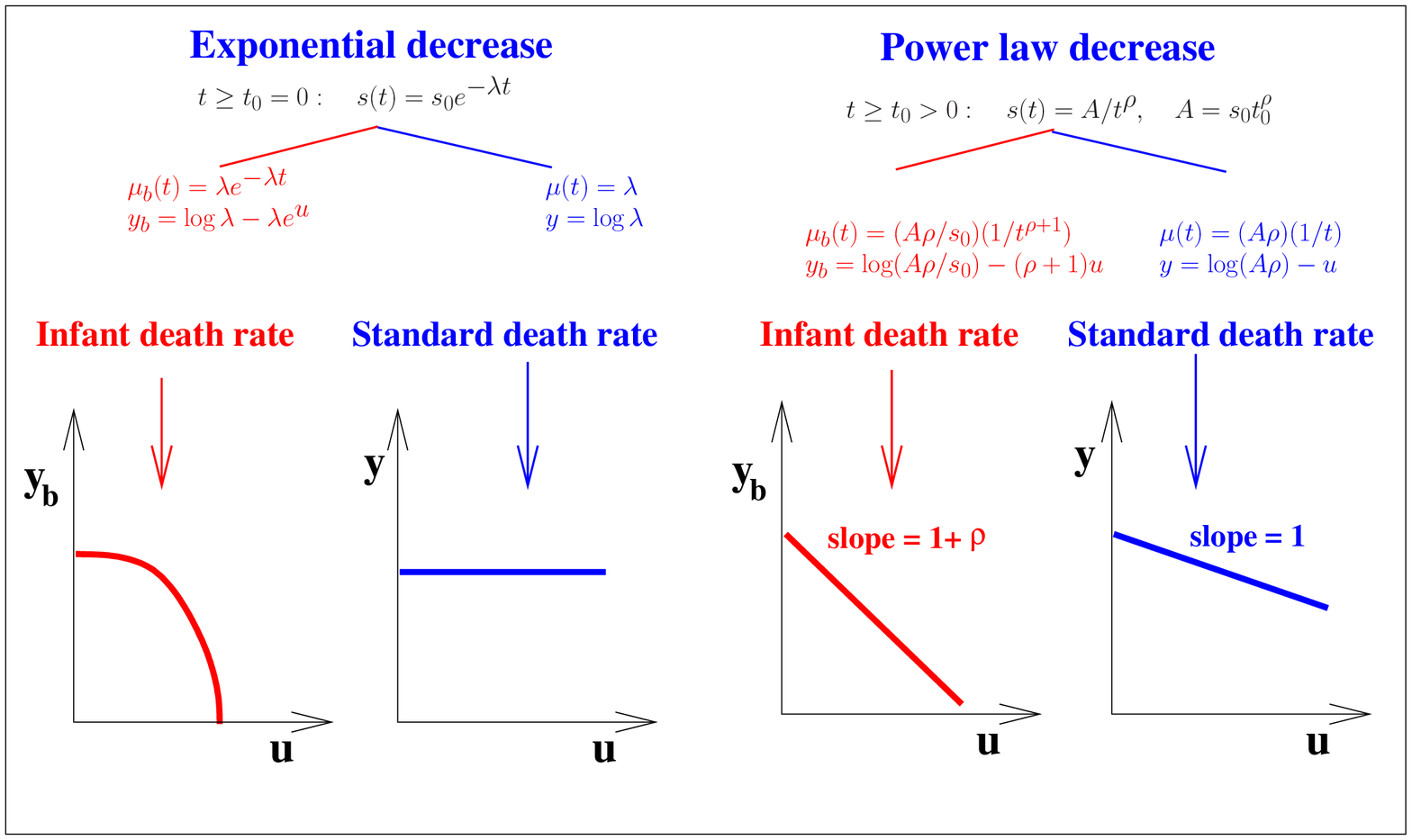}}
\qleg{Fig.\qhu A1b\qhv Log of infant death rate ($ y_b=\log \mu_b $) and
of standard death rate ($ y=\log \mu $) as a function of $ u=\log t $.}
{The shapes are shown in two typical cases: exponential and
power law decrease of $ s(t) $.
As expected, $ \mu(t) $ falls off slower than $ \mu_b (t) $
for the simple reason that for any $ t>0 $, 
the denominator $ s(t) $ is smaller than $ s_0 $.}
{}
\end{figure}

For a power law, the curves of $ \mu $ and $ \mu_b $ have same shape
but different exponents. 
Intuitively, this is quite understandable
because for a slow decrease of $ s(t) $ one has: $ \mu_b \simeq \mu $; 
in most cases examined in this paper $ \mu_b\sim 1/t $ which means
that $ rho $ is close to zero and $ s(t) $ almost level
during the infant mortality phase.
\qpar

In the next subsection we explain that this dual definition
is a consequence of how the data are recorded.

\qA{Reason of the dual definition of mortality rates}

The infant mortality definition is usually used for newborns
under one year. We have seen that the infant mortality
phase (i.e. falling mortality) in fact extends until 10 years.
So, why was a one year threshold selected?
\qpar

The (standard) age-specific death rate is a ratio of two numbers:
the numerator is the number of deaths which occurred in one
year in a given age-group (e.g. 15-19)
whereas the denominator is the number of (living) people 
in this age-group. The first number is provided by the death
certificates while the second is provided by the census.
\qpar

Now, let us try to compute the death rate of newborn babies
between 1 and 2 months of age with the previous
definition. There is no problem for the numerator because
death certificates give the date and time of 
birth and death which defines the age at death
with an accuracy of up to one hour.
For the denominator one would need the
number of living newborns of that age. 
However, this number will have strong monthly fluctuations
because of the seasonal variability of the birth rate.
In other words,
this definition cannot be used and must therefore be replaced
by another. In order to get a rate
the simplest way is to divide the deaths by the total number of
live births in the same year.
\qpar

Over one year of age one can in principle use
age-group census data but it should be observed that, 
specially for
young ages, the definition of an age group requires
a number of conventions. For instance, in order to compute
ages one must define
(somewhat arbitrarily)
a specific census date even though census operations may
take several months.  This shows that the infant death
rate is a simpler and clearer concept than the standard 
death rate. 
\qpar

These definitions have another noteworthy implication.
In the computation of an infant death rate one needs to
take into account the length of the age interval under
consideration. Thus, to get the rate per day for the 
age 1-2 months one will divide the number of deaths not only
by the live births but also by the length of the interval
expressed in days (i.e. 30 days). On the contrary, in computing the 
(standard) death rate for the 14-19 age group the
length of the interval (i.e. 5 years)
is irrelevant because the 
denominator refers to the same age group.

\qA{In search of ``natural'' infant mortality rates}

For humans as well as other living organisms infant death rates
may be inflated by temporary diseases or epidemics.
In addition, for species reared in laboratories or zoos
death rates may be amplified because the living conditions
are not good enough. For animals kept in zoos this can just
be the result of a lack of space, whereas for small species
(e.g. fish larvae) it may be due to the fact that we do not
well know how to feed them. \qL
On the other hand, especially in past decades, human
infant mortality has been
drastically reduced thanks to medical progress. This led to
the survival of babies afflicted by malformations who
would not have been viable otherwise. 
\qpar

This leads to the following practical rules for data
selection.
\qbu We discard all data collected in the field that is to say
in conditions (e.g. bad weather, predators) which cannot be
controlled.
\qbu We discard the data recorded in populations
in which the spread of a
disease or other inappropriate conditions
have led to high mortality rates. For instance,
there are recurrent disease outbreaks in hatcheries, fisheries 
or zoos.

\qA{Possible data sources}

For reasons which are easy to understand the process
of aging gets considerable attention. 
Within the broader discipline of biodemography, gerontology
has become a field in itself with its own journals
(e.g. ``Experimental Gerontology'' or ``Biogerontology'').
Yet, as observed by Levitis et al. (2013), ``gerontologists
focus on aging and usually take neither data on, or interest in,
the periods before adulthood''. Apart from the gerontologists.
life tables are occasionally set up by ecologists.
As an example, one can cite the work of It\^o (1980). We already
mentioned the fact that almost none of the life tables
contained in his book can be used for studying infant mortality.
In this respect the key-requirement is to start the observations as
shortly as possible after birth or germination.
If the first data point refers to
a few weeks after birth (as is seen in many survivorship curves)
the main part of the information will be lost. 

\qA{Suggested experiments}

Although the data reported in the present paper cover
a wide range of species, they are all complex multicellular
organisms. One would need empirical evidence for 
``simpler'' organisms.
What organisms would be particularly appropriate?
\qpar

Because of their relative simplicity, it is tempting
to use  unicellular organisms.
In some cases such as
{\it Euglena gracilis}, a unicellular swimming 
organism about 50 micrometers in length, reproduction
leads to clones of the parent organism. In other cases,
such as {\it Paramecium caudatum} a swimming organism
about 100 micrometers in length,
the reproduction by
division can lead to offsprings which are not identical
to the parent (through the process of an exchange
of genetic material between two organisms prior to division).
However, no matter whether the children are clones or not,
there is the challenge of 
distinguishing those which died from
those alive. This, in different form, is the difficulty
already discussed for plants. Unfortunately, even 
for phototropic organisms (such as {\it Euglena gracilis}) 
there is always a small percentage who do not respond to the
stimulus which means that this method can hardly
be used.
\qpar

{\it Caenorhabditis elegans}, a little worm about
1mm long which has only about one thousand cells, would seem
to be a good candidate. The adults lay eggs which
hatch after a few hours. The death rate in the 12 hours
following hatching is about 4\% per day; it falls to
about 1.5\% per day in the following 16 hours
(Smith 2011 p. 24, Tew 2008 p. 26). At sexual maturity
it is at a minimum level of 1\% from which it
then increases exponentially over a period of
30 days at the end of which it reaches 100\%.
\qpar

We hope that the publication of the present paper will
lead other researchers to perform
accurate measurements of infant
mortality rates. We are confident that 
once more observations become available covering
a wide range of cases a clearer understanding will emerge.

\qI{Appendix B. Indications about possible models}

As already noted, at this point of our investigation
it is too early to propose a full fledged model.
However, it may be useful to give some hints in order
to start the discussion.
\qpar
The models delineated below are able to generate a power law
under some more or less plausible assumptions.
However, to be honest, one must recognize that there are
probably many alternative models which can 
also generate power laws under
``plausible'' assumptions, especially since plausibility
is a fairly subjective notion. A better criterion is to
ask whether or not the models have a predictive power.

\qA{Models based on interdependent defects}

We assume that at birth a living organism $ i $ 
has a number of defects, $ F_i $ which determines its 
probability of dying.
For instance, one may say: ``An individual with 20 defects will die
in one day, whereas one with only 5 defects will die in one week''.
For a population cohort of $ n $ individuals
the average number of defects will be: 
$ {\overline F}(t)=(1/n)\sum_1^N F_i(t) $ and the assumption made
at individual level will translate into
a relationship between $ \overline F $ and
the age-specific frequency of death in the population,
that is to say the (normalized) age-specific death rate $ \mu_b(t) $.
Through this assumption the survival
problem is reduced to a discussion regarding defect numbers.
A simple form of the functional relationship
would be proportionality: $ \mu_b(t)=k{\overline F}(t) $;
this is what will be assumed thereafter but
any increasing function would be acceptable as well.
\qpar

The assumption that the death rate is proportional
to the average number of defects is of course fairly natural, but 
now we must also explain why the death rate decreases fastest
when it is highest (namely immediately after birth) and 
then slower as individuals become older.
\qpar

There are three mechanisms through which the
average number of defects may decrease in the course of time.
\qee{1} Through the death of individual $ i $ the term $ F_i $
will be reduced by zero. As the first individuals to die will 
be those with the largest numbers of defects the decrease
of the average will be fastest immediately after birth.
The individuals who die after a while may have
only one or two defects so their deaths will reduce
$ \overline F $ very little.
\qee{2} There will also be a self-correction mechanism.
For instance, a major source of defects is birth prematurity
but for those newborns who remain alive, their defects will correct
themselves as they become older just because of natural
maturation.
\qee{3} An additional effect may concentrate the deaths
shortly after birth, namely that fact 
that the defects are certainly not independent.
Incidentally, the interdependence of defects may
be experienced 
when executing computer code.
A single mistake, e.g. a variable that was
not defined, has a cascading effect and
results in an avalanche or error messages. Once this
mistake is corrected, everything falls in order.
The same effect
may happen in living organisms.
For instance,
suppose that for some reason an organ $ O $
which produces an hormone $ H $ is underdeveloped
because enzyme $ E $ is in low
supply. So, at this stage, there are 3 interdependent defects:
low level of $ E $, underdeveloped organ $ O $ and low level
of $ H $.
Then, by correcting the core defect $ E $ 
the two other defects will fall in line: $ O $ will
reach the right size and the concentration of $ H $ will
reach the right level. \qpar
It is reasonable to assume that defect interdependence is 
highest when $ F_i $ is largest. Thus, self-correction will lead
to a very fast decrease of $ \overline F $ at the beginning
as long as individuals with many defects are present.
\qpar
An argument which gives credence to such a mechanism is the close
correlation that is observed
between low weight and premature birth.
Instead of speaking of defect numbers, this mechanism
could be rephrased  in terms of unfinished growth. 
If the premature baby
is kept in an incubator and fed appropriately, it will
be able to terminate its growth process. For instance,
the lungs, one of the last organs to be
completed, will eventually function properly with
the consequence that
the appropriate level of oxygen will be delivered 
in the blood flow. This will {\it ipso facto}
result in the improved working of many organs
across the whole body.
Naturally,
this argument does not apply to
malformations which cannot be self-repaired. 
\qpar

Before closing the discussion of this model
one can propose a plausible
conjecture. \qL
If the system has many initial causes of failure
corresponding to various mechanisms it
seems reasonable to expect that the dispersion
in their time constants will increase along with their number.
The infant mortality phase comes to an end when all initial defects
have emerged and have been eliminated either through
a self-repairing process or by elimination of the
faulty items.
Thus, one would expect
the length of the infant mortality phase to increase
with the number and diversity of the defects,
that is to say basically with the ``complexity'' of
the organism.\qL
Note that in order to apply this argument
to living organisms one must renormalize the time scale
with respect to size because, roughly
speaking, the life span of living 
organisms increases with their size.
\qpar

We now turn to a class of models which relies on a global
parameter, for instance the birthweight of the organism
or more generally any parameter of significance for the death
rate.

\qA{Model based on the incidence of birthweight}

In this subsection we will ask ourselves how
a power law distribution might
arise from a Gaussian distribution due to a simple heterogeneity 
in an important physical parameter.
\qpar

In humans
low birthweight is a crucial determinant of infant mortality.
Here, however we are not interested in global infant mortality
but in the age-specific pattern of infant mortality. 
So far, in order to describe this pattern we have used the
infant mortality rate $ \mu_b(t) $; here, however, we wish
to describe it by a density probability function. For instance
the density function of the the life span $ T $ (seen as a random
variable) would be:
$$ f_T(t)dt ={ \hbox{\small Number of individuals who die in the
age interval } (t,t+dt) \over 
\hbox{\small Total number of individuals} } $$

The right-hand side is nothing else than $ \mu_b(t)dt $
(except for a possible normalization factor which is 
unimportant here); thus: $ f_T(t)\sim \mu_b(t) $.
\qpar

The next step is
to establish a connection between the distribution of
birthweights and the distribution of life spans. 
For instance, one may say: ``For a birthweight $ W $ in the interval
(1.5kg,1.6 kg) the life duration $ T $ will be about 
one week (on average), whereas
for a birthweight in the range (2.0kg,2.1kg) it will be 4 months''.
More generally, for the two random variables $ W $ and $ T $ 
one may posit a relationship of the form: $ T=g(W) $. Clearly,
$ g(w) $ should be an increasing function, and in addition
one should have $ g(0)=0 $. A possible example would
be a parabolic relationship: $ T=W^2 $.
\qpar

Now, how can we get a power law density function for $ T $? \qL
The distribution of $ W $ is known to be Gaussian%
\qfoot{Yet the tails are somewhat more heavy than would
be expected from a standard Gaussian. That is why it was suggested
by some authors to represent it by a mixture of two Gaussian
distributions: one for the
central values and a second one, much flatter (that is to say
with a large standard deviation), which would account for the tails.}%
, 
which
means it has the following density function%
\qfoot{Of course, $ W $ takes only positive values which
means that $ f_W(w) $ should be zero for $ w \le 0 $.}%
:
 $$ f_W(w)= { 1 \over \sigma \sqrt{2\pi} }
\exp \left[-(w-m)^2/2\sigma^2 \right] \qn{B.1} $$ 

What will be the distribution of $ T $?\qL
According to a well-known result,
if the equation $ t=g(w) $ has only one solution $ w_1=g^{-1}(t) $
the density function of $ T =g(W) $  will be (Papoulis 1965, p. 126):
 $$ f_T(t)= { f_W\left(w_1(t) \right) \over g'\left(w_1(t) \right) }
  \qn{B.2} $$
where $ g'(w) $ denotes the derivative of $ g(w) $.
\qpar

As an illustration, we consider the case: $ T=g(w)=w^{\alpha} $.
It leads to: $ g'(w)=\alpha w^{\alpha -1} $ and 
$ w_1(t)=t^{1/\alpha} $. 
Clearly the power law behavior
can only come from the denominator, not from the numerator.
Thus, we must have:
 $$ \left( t^{1/\alpha} \right)^{\alpha-1} \sim t^{\gamma} $$

which leads to: $ \alpha=1/(1-\gamma) $ that is to say 
$ \gamma =1-1/\alpha $.
\qpar

 The resulting power law behavior is little
affected by the function which stands at the numerator
provided that $ (w_1-m)/\sigma $ is not too large; otherwise the
exponential becomes so small that everything else is suppressed.
\qpar

Fig. B1a,b shows the case $ \alpha=4,\ \gamma=0.75 $. The graph
on the left-hand side shows the functions (``g'' means ``Gaussian''
and ``h'' means ``hyperbolic'')
$$ f_T(t)\sim y_gy_h\quad  y_g=\exp\left[-(t^{1/4}-m)^2\right]/(2\sigma^2),
\quad y_h=1/t^{\gamma} $$
The graph on the right-hand side shows the results of a simulation
with the same parameters.
\qpar

{\bf Remark}\quad Why has the transformation $ T=W^4 $ 
(the exponent 4 is just for the purpose of illustration,
the argument is the same for any exponent larger than 1)
the
dramatic effect of making $ f_T(t) $ diverge at $ t=0 $?
At first sight, this transformation seems fairly smooth.
Yet, it has one dramatic feature: the derivative of the
inverse function $ g^{-1}(t)=t^{1/4} $ diverges at $ t=0 $.
The reason why this leads to the divergence of $ f_T(t) $
is easy to explain intuitively.
\qbu Firstly, the fact that the 
initial distribution is Gaussian is unimportant.
For the sake of simplicity we can as well assume that
$ f_W $ is a uniform distribution that is to say has a
rectangular density function.
\qbu The probability 
assigned to the interval $ t\in I_1=(0.01,0.101) $ 
will be the same as the probability assigned to
the interval $ w\in J_1=\left( 0.01^{1/4},0.1^{1/4} \right)=
\left( 0.32,0.56 \right) $. Similarly, the
probability assigned to $ t\in I_2=(15.0,15.1) $ will be given
by the probability of $ w\in J_2= \left( 15.0^{1/4},15.1^{1/4} \right) 
=\left( 1.968,1.971 \right) $. Thus, for $ I_1, I_2 $ of the
same length, $ J_1 $ is 80 times larger than $ J_2 $.
As the distribution of $ W $ is supposed uniform, we will have
$ P\{t\in I_1\}=80P\{t\in I_2\} $. \qL
The same argument is illustrated
in the insert graph of Fig. B1 b. The intervals $ I_1,I_2 $
were selected for the sake of making the graph best readable:
$ I_1=(0.2,2.5),\ I_2=(15.17.3) $.
With these intervals $ J_1 $ is 7.5 times larger than $ J_2 $.
\qL
%
\begin{figure}[htb]
\centerline{\psfig{width=17cm,figure=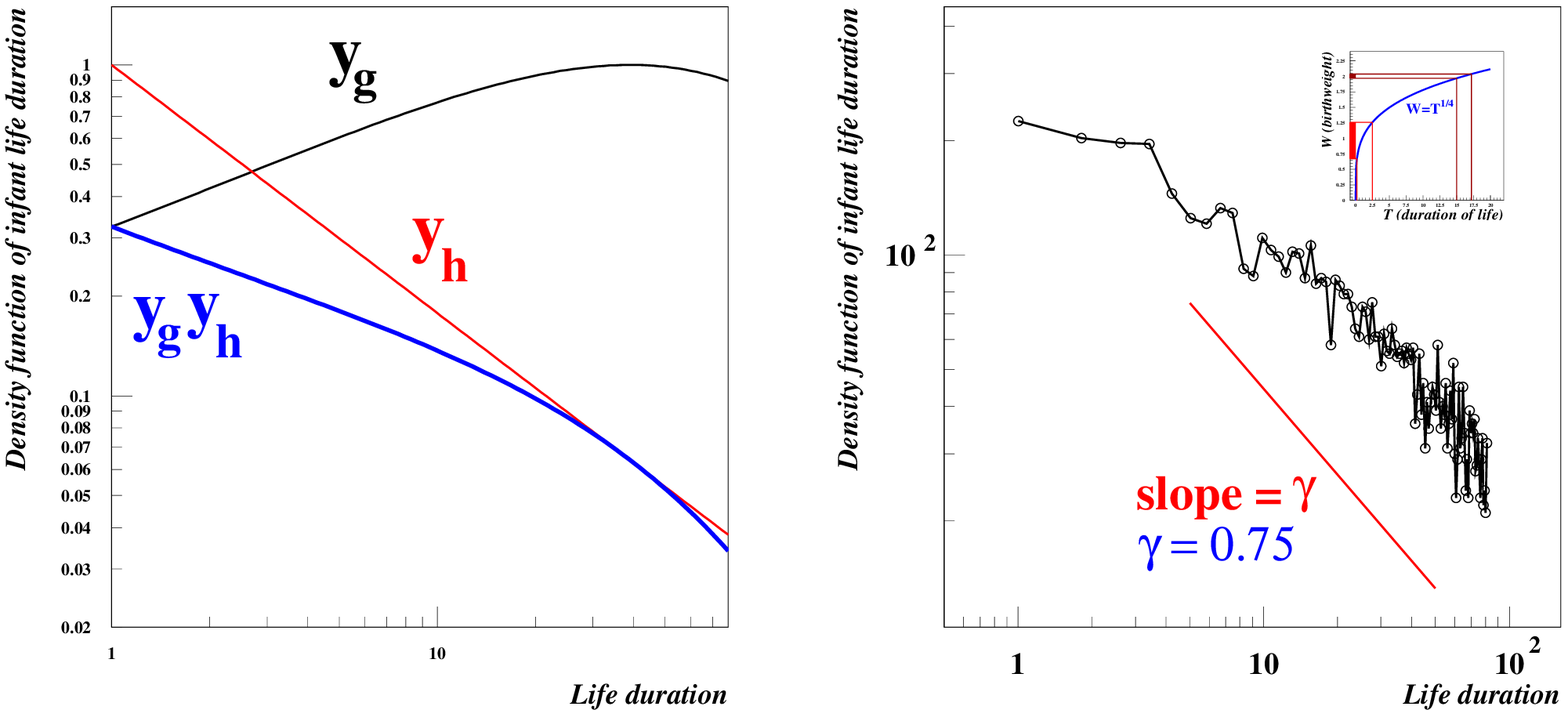}}
\qleg{Fig.\qhu B1a,b\qhv Density function of life duration: 
left: calculated from the theoretical expressions, right:
generated through a random number simulation.}
{The Gaussian distribution describing the weight distribution has
the parameters $ m=2.5,\ \sigma=1 $. The simulation
relied on $ 10,000 $ random drawings. The insert gives an
intuitive explanation of 
why the transformation $ T=W^4 $ leads to a density function
$ f_T $ which has a very high probability in the vicinity
of $ T=0 $.}
{}
\end{figure}
Incidentally, the requirement of the conservation of the probability
mass that we used above also quickly
leads to the result (2). Indeed:
$$  f_T(t)dt=f_W(w)dw \Rightarrow f_T(t)=f_W(w){ dw\over dt }
=f_W(w){ dg^{-1}\over dt }=f_W(w_1){ 1\over g'(w_1) } $$
\qpar

Are such high exponents consistent with what is known about
the  relationship $ T=g(W) $? There are three answers to
this question.
\qee{1} So far we did not find any data giving the function
$ T=g(W) $. In medical statistics it is another variable
which is commonly used namely the ``birthweight specific infant
mortality'' which means infant mortality as a function of 
birthweight.
In this expression, ``infant mortality'' has
the standard medical meaning of ``under one year of age''
rather than the broad meaning used in the present paper.
\qee{2} As a function of birthweight, infant mortality increases
sharply when $ W $ decreases. For $ W=2.0 $ kg, the infant mortality
is 20\%, then $ W=1.5 $ kg gives a mortality of 40\% and $ W=1 $ kg 
of 80\%.
In a broad sense, such a rapid change is consistent with
exponents $ \alpha $ that reach high values.
\qee{3} At first sight it might seem that in
the present model we cannot get $ \gamma>1 $.
This is not really true however as can be seen in 
Fig. B1a. It shows that the slope of $ f_T(t)=y_gy_h $ is
strictly equal to $\gamma $ only in the vicinity
of the maximum of the function $ y_g $. 
On the left-hand side of the
maximum the slope will be smaller than $ \gamma $ whereas
on the right-hand side it will be higher than $ \gamma $. 
\qee{4} In the previous argument we considered $ T $ as
a function of birthweight but this is only one
possible case. $ T $ can be connected as well to
other variables, for instance to the effectiveness of
the immune system at birth. The main reason for selecting
the birthweight was data availability. The effectiveness
of the immune system
may be a ``better'' variable but,
to our best knowledge, no data are available. 

\vfill\eject

\qA{How can one generate a power law death rate by superposition?}

The transition from a specific death rate (e.g. exponential
or Gaussian)
to a 
power law death rate can be presented in a more
general framework. 
\qpar

Let the death rate be conditional on a parameter $ a $,
that is to say: $ \mu\equiv \mu(t|a) $ and suppose
that the parameter $ a $ is determined by a distribution $ f(a) $.
Under these assumptions the unconditional death rate 
resulting from the superposition of the processes 
$ \mu(t|a) $ is:
$$ \mu(t)=\int_0^{\infty} \mu(t|a)f(a) da \qn{B.3} $$

The expression (B.3) is very general. It gives the
mortality rate  that results from the superposition of 
a set of ``elementary'' processes with weights $ f(a) $.
\qpar

As an illustration,
let us consider the following case:
$$ \mu(t|a)=Ae^{-at},\quad 
f(a)=\left[{1 \over b\Gamma(d)}\right]\left({a \over b}\right)^{d-1}
e^{a/b} \qn{B.4} $$

The function $ f(a) $ is the density function of a Chi-squared 
distribution with $ d $ degrees of freedom. With this
choice we obtain:
 $$ \mu(t)={ A \over (1+bt)^d } \qn{B.5} $$

In order to connect this expression with observation 
we write it in log-log form by defining $ y=\log\mu $ and
$ u=\log t $:
$$ y=-d\log (1+be^u)+\log A \qn{B.6} $$

For large values of age%
\qfoot{In the case of human mortality, ``large'' would
mean a few years.}%
, 
typically $ t\gg 1/b $, one gets: $ y=-du+\log (A/b^d) $ which
leads to the identification: $ d=\gamma $. 
\qpar

For $ t\ll 1/b $, one gets: $ y=-dbe^u + \log A $ which shows
a behavior similar to what was observed for beetles (Fig. 11).
\qpar
For this model it would be an important step forward to be
able to identify the parameter $ a $ with a physical
variable for this will 
provide a more direct contact with observation.

\count101=0  \ifnum\count101=1
For small $ t $ observation shows that $ \mu(t)\sim 1/t^{\gamma} $.
Can we get this behavior if we keep $ \mu(t|a)=Ae^{-at} $?
With this choice, (B.3) can be interpreted as a Laplace transform.
For such a transform one knows that:
 $$ \lim_{a\rightarrow \infty}f(a) = \lim_{t\rightarrow 0}t\mu(t) \qn{B.6}$$ 
 
With the $ \mu(t) $ given by observation the limit 
of the right-hand side  is $ 0 $ if $ \gamma<1 $,
$ 1 $ if $ \gamma=1 $ and $ \infty $ if $ \gamma>1 $.
This gives indications for the selection of $ f(a) $.
$ f(a) $ can be interpreted as a 
probability density function over $ (0,\infty ) $ only
in the first case, but in reality $ a $ will certainly be
limited to a finite interval $ (0,a') $ which means that $ f(a) $
can always be interpreted as a density function that would be
equal to zero for $ a>a' $.
\fi

\vfill\eject

\qA{Comparison of the two classes of models}

The adaptive mechanism on which the first model relies
establishes a connection
between what we observe (i.e. the age-specific infant mortality)
and what is going on inside the system. This could give an insight
and understanding of the phenomenon of infant mortality. However,
as already pointed out, because of its many free parameters
the model has no predictive power. This obstacle may be overcome by
doing additional observations focused on simple subsystems.
\qpar

Unlike the first, the second kind of models
does not relate what is observed to what is happening
inside. Yet, it has the merit of offering testable predictions.
The successive steps can be summarized as follows:
$$ \hbox{\normalsize Measurement of }\gamma \rightarrow \alpha
\rightarrow T=g(W) \rightarrow 
\hbox{\normalsize Comparison of } g(W)
\hbox{\normalsize to observation} $$

If the predicted form of $ T=g(W) $ agrees with observation,
we will get more confidence in the model, yet we will not learn anything
new. On the contrary, if the predicted form does
{\it not} agree with observation, this will raise a question.
If $ T=g(W) $ displays a well defined pattern but which differs
from the predicted function, then one must understand why.

\qI{Appendix C: Additional data about plant mortality}

Is the high exponent found for palm trees also observed
for other big trees? As already said,
there are few studies available which is why their discussion
is made in the present appendix rather than in the paper itself.
One study that is cited in Valen (1975),
namely Hett and Loucks (1971), concerns a species of maple
tree. Although less high than {\it Euterpa}, it is also a big tree
which can live up to 250 years. Observations were conducted
in Wisconsin for trees ranging from germination time to the age of
16 years and they led to $ \gamma=1.35\pm 0.16 $. This value
shows that the $ \gamma $ value of {\it Euterpa } may be exceptional%
\qfoot{Although Valen cites the paper by Hett and Loucks,
he does not discuss their results and why they are so different;
in fact, he does not even give their results.
This is just a
confirmation of the fact that, unlike
physics, life sciences and social sciences develop 
a kind of ``mosaic knowledge''.
By this term we mean
that basically these fields are made up of a multitude of disconnected
results.
This can be illustrated by the title of
a randomly selected paper: 
``The demography of the short-lived perennial halophyte {\it Spergularia
maritima} in a sea-shore meadow in south-western Sweden''
by A. Telenius, Journal of Ecology 81,1,61-73 (1993).
The study concerns {\it one species} in {\it one place},
in {\it one country}  
without any attempt
to broaden it to other places or other species.}%
.
\qpar

Before closing this subsection about plants a word of caution
must be added to observe that the methodology used for trees
relies on the assumption that the age structure of the
population under observation is in a stationary state. Why?
The reason is simple. Valen's observations cover an age
interval of 150 years. Clearly such an observation
was not made by following a cohort of seedlings in the course
of time. Instead Valen conducted a transversal study of the
population. Although the study by Hett and Loucks covers only
16 years of age, they used the same method. One should
keep in mind that,
due to possible fluctuations in the age structure,
transversal studies may not correctly reflect and describe
the survivorship curve. In other words, the values
of $ \gamma $ need to be confirmed by plant studies
based on longitudinal analysis.
\qpar

One study of this kind can be found in Silvertown and Dickie
(1980). They followed the growth of several perennial
plants (that is to say plants which live several years)
after their germination over a period of one year. 
On average the authors made a field visit every 42 days.
For several of their  species, their initial cohort comprised 
less than one hundred plants.
We will limit ourselves to one species (namely {\it Anthyllis
vulneraria}, a plant about 30 cm high with green leaves
and yellow flowers)
whose initial
cohort numbered over 300. 
The values of the infant mortality death rates derived from the
survivorship curve plotted in the paper
lead to the following exponent: $ \gamma=3.4\pm 1.9 $.
This high exponent suggests a high initial death rate 
and the latter
is quite consistent with the qualitative
observations reported by the authors.
They tell us that ``no individuals reached the
flowering stage in any of the population under observation
and most cohorts experienced over 80\% mortality in their
first year''. They add that such high levels of pre-reproductive
mortality agree well with the observations made by other researchers.
In the particular case of {\it Anthyllis vulneraria}, a study
conducted  in the Netherlands on 1,117 plants showed that only
27\% ever reached the flowering stage (Sterk 1975).
Moreover, the author notes (p. 333) that 
``younger plants have a smaller
survival chance owing to the high seedling mortality''.

\qA{Observations in greenhouse under controlled conditions}

The exponents $ \gamma $ observed so far were equal to 
$ 2.6,\ 1.3,\ 3.4 $ respectively
which gives an average of $ 2.43 $ that is to say
two or three times higher than the exponents found for animal
species. As these were field observations
should these high values not be attributed to
the fact that in the field the
trees are exposed to various
adverse conditions? In other words
can these observations be confirmed
by controlled observations performed in greenhouses?
\qpar
We will mention one observation of that kind.
Lidia Cruz and her collaborators (2014) report an
experiment in which the seedlings of 6 cacti species were
grown in a greenhouse during a period of 16 months
after germination.
For the species {\it E. reichenbachii} the exponent was 
$ \gamma=2.6 $ and the average for the 6 species was 1.98.
This shows that high values of $ \gamma $ 
are not limited to field observations of trees.
However, not all trees have a $ \gamma $ higher than 1.
An experiment for acacia trees performed in India  
gave values of $ \gamma $ around 0.7 (Shaukat et al. 1999).

\vskip 5mm

{\bf Acknowledgments}\quad We would like to thank
Prof. Thomas Kirkwood for his
encouragements. His paper on the Gompertz law (Kirkwood 2015)
was a source of inspiration and a guide in writing
the present paper,

\vskip 5mm

{\large \bf References}
\vskip 2mm

{\color{blue} 
The objective of the comments within square brackets
is to indicate the implications of the work under consideration
for the present investigation. They may be removed
in the final version.}
\vskip 2mm

\qparr
Berger (Y.M.) 1997: Lamb mortality and causes. A nine-year summary at
the Spooner Agricultural Research Station. In: Proceedings of the 45th
Annual Spooner Sheep Day,
University of Wisconsin-Madison, pp. 30-41.

\qparr
Bianconi (E.), Piovesan (A.), Facchin (F.), Beraudi (A.), 
Casadei (R.), Frabetti (F.),
Vitale (L.), Pelleri (M.C.), Tassani (S.), 
Piva (F.), Perez-Amodio (S.), Strippoli (P.),
Canaider (S.) 2013: 
An estimation of the number of cells in the human body.
Annals of Human Biology 40,6,463-471.\qL
[The estimate found by the authors is $ 3.7\times 10^{13} $.]

\qparr
Buffenstein (R.) 2008: Negligible senescence in the longest
living rodent, the naked mole-rat: insights from
a successful aging species. 
Journal of Comparative Physiology B 178,439-445.

\qparr
Campbell (N.A.), Reece (J.B.) 2004: Biologie. Brussels, De Boeck.

\qparr
Cousin (X.), Richmond (P.), Roehner (B.M.) 2016:
Unraveling infant mortality: the case of zebrafish.
Preprint February 2016.

\qparr
Cruz (L.R.S.), Pournavab (R.F.), Jimenez (L.D.), Hernandez-Pinero,
Parra (A.C.), Avila (M.L.C.) 2014: Seed germination and seedling
survival of 6 cacti species using natural zeolite as substrate,
International Journal of Current Research and Academic Review
2,9,81-91.

\qparr
Espinas (A.) 1878, 1935: Des soci\'et\'es animales
[About animal societies].
Thesis, University of Paris. In 1879 the thesis was
translated into German and in 1882 into Russian.
It does not seem to have ever been translated into English.
In 1935, it was re-edited by ``Librairie F\'elix Alcan'', Paris.\qL
[This study was written in a time marked by the emergence
of sociology as a new field. It is interesting to observe
that at that time it seemed natural for the author to
discuss human and animal societies within the same 
conceptual framework. Nevertheless, Alfred Espinas 
is quite aware of the fact that
in order to be meaningful such broad comparisons
must focus on a small number of basic 
and well selected features. \qL
The author explains his epistemological positions in
the following terms (p. 5), followed by my translation:\qL
``Nul \^etre vivant n'est seul. La vie en commun n'est pas dans
le r\`egne animal un fait accidentel. Elle n'est point,
comme on le croit souvent, le privil\`ege de quelques esp\`eces 
isol\'ees dans l'\'echelle zoologique, castors, abeilles
et fourmis; elle est au contraire un fait normal,
constant, universel.\qL
C'est une tentative aussi vaine que fr\'equemment renouvel\'ee
que celle de d\'e\-cou\-vrir les lois de la vie sociale dans 
l'homme ind\'ependamment de toute comparaison avec les autres
manifestations  de la vie sociale dans le reste
de la nature. Mais il faut reconna\^{\i}tre qu'un simple
rapprochement ne suffit pas''. \qL
No living organism is completely isolated. 
The act of living together
is not limited to a small number of social species such as
beavers, bees or ants. Quite on the contrary it is a universal
feature of animal life. It would be a vain and worthless attempt
to try to unravel the laws of social life independently of the
study of social behavior in animals. Yet, one must recognize that 
in order to be fruitful such a comparison must be well focused.]

\qparr
Garrido (S.), Ben-Hamadou (R.), Santos (A.M.P.), Ferreira (S.),
Teodosio (M.A.), Cotano (U.), Irigoien (X.), Peck (M.A.), 
Saiz (E.), R\'e (P.) 2015: Born small, die young: intrinsic,
size-selective mortality in marine larval fish.
Scientific Reports (published online on 24 November 2015).

\qparr
Gavrilov (L.A.), Gavrilova (N.S.) 2006: Models of system failure
in aging. In: Conn (P.M.) editor: Handbook of models for
human aging, p. 45-67.  Elsevier.\qL
[The authors describe several mathematical structures
which {\it may} play a role in the process of aging,
for instance systems working in series (i.e. fast global failure)
or in parallel (i.e. delayed global failure),
systems beset with initial damages, avalanche-like situations where
new defects appear in proportion to the number of defects 
already existing. Based on these effects (or their combination)
it is possible to generate Gompertz's law of aging in many
ways. However, unless confirmed by the observation 
the internal structure, such models remain merely a mathematical game.
Better than an avalanche of speculative models largely
based on thin air, modeling should be done hand in hand
with appropriate experiments
One should not think that such a situation where several
theoretical frameworks are in competition
is special to the life sciences. 
It also occurs in physics.
For instance, the solubility of gases in water can be
explained by several effects.
\qbu Equality at equilibrium of the flows in both directions.
As the velocities are known from the equipartition
principle, it is possible to derive the concentrations.
\qbu  Number of molecules of water in
the solvation shell, i.e. the shell surrounding a molecule
of gas (Sharlin et al. 1998)
\qbu Concentration gradient in the film at the
interface between the gas and the liquid (Lewis and Whitman 1924).
\qbu Respective
interaction strengths between molecules in the gas on the one hand
and in the liquid on the other hand. Obviously, if the molecules 
of the gas and of the liquid have a strong attraction solubility
will be higher.

\qparr
Gisbert (E.), Williot (P.), Castell\'o-Orvay (F.) 2000:
Influence of egg size on growth and survival of early
stages of Siberian sturgeon ({\it Acipenser baeri}) under
small scale hatchery conditions. 
[Each experiment involved about 3,000 eggs obtained 
from twenty 13-14 year old females and it was
repeated several times.]

\qparr
Glastonbury (J.R.W.) 1976: A survey of preweaning mortality
in the pig. Department of Agriculture, Wollongbar, New South
Wales, Australia.

\qparr
Grove (R.D.), Hetzel (A.M.) 1968:  Vital statistics rates 
in the United States, 1940-1960. United States Printing Office,
Washington DC. 

\qparr
Hanley (M.E.), M. Fenner (M.), Whibley (H.), Darvill (B.) 2004:
Early plant growth: identifying the end point of the seedling
phase. New Phytologist 163,61-66.

\qparr
Hett (J.M.), Loucks (O.L.) 1971: Sugar maple ({\it Acer Saccharum} Marsh.)
seedling mortality.
Journal of Ecology 59,2,507-520. 

\qparr
Hird (D.), Henrickson (R.), Hendricks (A.) 1975:
Infant mortality in {\it Macaca mulatta}: neonatal and postnatal
mortality at the California Primate Research Center.
Journal of Medical Primatology 4,8-22.

\qparr
It\^o (A.) 1980: Comparative ecology. Cambridge University
Press. \qL
[The first Japanese edition of this study [{\it Hikaku Seitaigaku}]
was published in 1960.
In chapter 2 the author gives many examples of survivorship
curves for insects, fishes, reptiles, birds, mammals. 
Needless to say, it is not easy to establish meaningful
comparative
links between species which are as different as Atlantic
mackerels and African elephants; the mackerel female lays
almost one million eggs whereas the elephant female has only
one offspring at a time.]

\qparr
Kirkwood (T.B.L.) 2015: Deciphering death: a commentary
on Gompertz (1825) ``On the nature of the function
expressive of the law of human mortality, and on a new
mode of determining the value of life contingencies''.
Philosophical Transactions of the Royal Society B
(6 March 2015).

\qparr
Kohler (I.V.), Preston (S.H.), Lackey (L.B.) 2006: 
Comparative mortality levels among selected species of
captive animals. Demographic Research 15,4,413-434.

\qparr
Laskar (R.) 1970: Feeding, growth, and survival of 
{\it Engraulis mordax} larvae reared in laboratory.
Marine Biology 5,4,345-353.

\qparr
Laurel (B.J.), Brown (J.A.), Anderson (R.) 2001:
Behavior, growth and survival of redfish larvae in relation
to prey availability.
Journal of Fish Biology 59,884-901.

\qparr
Levitis (D.A.), Martinez (D.E.) 2013: The two-halves of
U-shaped mortality.
Frontiers in Genetics, vol. 4, article 31, 1-6. 

\qparr
Lewis (W.K.), Whitman (W.G.) 1924: Principles of gas absorption.
Industrial and Engineering Chemistry, December 1924,1215-1220.

\qparr
Mortality Statistics of 1910. Bulletin 109 published 
by the Bureau of the Census in 1912.
Death of infants from each cause, by days for the
first week of life, by weeks for the first month, and by months for
the first two years. Government Printing Office, Washington DC.

\qparr
Papoulis (A.P.) 1965: Probability, random variables and
stochastic processes. McGraw-Hill Kogakusha, Tokyo.

\qparr
Pearl (R.), Park (T.), Miner (J.R.) 1941: Life tables for
the flour beetle {\it Tribolium confusum} Duval. 
The American Naturalist 75,756,5-19.\qL
[This paper is the 16th study in a series entitled
``Experimental studies on the duration of life''.]

\qparr
Peleg (M.), Cole (M.B) 1998: Reinterpretation of microbial survival
curves.
Critical Reviews in Food Science and Nutrition 38,5,353-380.

\qparr
Price (R.A.), Anver (M.R.), Garcia (F.G.) Simian neonatology.
I. Gestional maturity and extrauterine viability.
Veterinary Pathology 9,301-309.

\qparr
Pouillard (V.) 2015: En captivit\'e. 
Vies animales et politiques humaines dans les jardins zoologiques 
du XIXe si\`ecle \`a nos jours : 
m\'enagerie du Jardin des Plantes, zoos de Londres et Anvers.
[In captivity. Zoo management and animal lives in the
zoological gardens of Paris, London and Antwerp
from the 19th century to 2014.]. PhD thesis.
Universit\'e libre de Bruxelles and University of Lyon 3.

\qparr
Richmond (P.), Roehner (B.M.) 2016a:  Predictive implications
of Gompertz's law.\qL
Physica A 447, 446-454.
Also available on the arXiv website 
at the following address:\qL
http://arxiv.org/abs/1509.07271

\qparr
Richmond (P.), Roehner (B.M.) 2016b: Effect of marital status
on death rates. Part I: High accuracy exploration of the
Farr-Bertillon effect. \qL
Physica A 748-767. 
Also available on the arXiv website at the
following address: \qL
http://lanl.arxiv.org/abs/1508.04939

\qparr
Richmond (P.), Roehner (B.M.) 2016c:  Effect of marital status
on death rates. Part II: Transient mortality spikes.\qL
Physica A 768-784. 
Also available on the arXiv website at the
following address: \qL
http://lanl.arxiv.org/abs/1508.04944

\qparr
Sahin (T.) 2001: Larval rearing of the Black Sea Turbot,
{\it Scophthalmus maximus} (Linnaeus, 1758), under
laboratory conditions.
Turkish Journal of Zoology 25,447-452.

\qparr
Scharlin (P.), Battino (R.), Silla (E.), Tu\~n\'on (I.),
Pascual-Ahuir (J.L.) 1998: Solubility of gases in water. 
Correlation between solubility and the number of water molecules
in the first solvation shell.
International Union of Pure and Applied Chemistry (IUPAC)
70,1895-1904. Paper presented at the 8th International 
Symposium on Solubility Phenomena, 5-8 August 1998, Niigata, Japan.

\qparr
Shaughnessy (P.W.), DiGiacomo (R.F.), Martin (D.P.), Valerio (D.A.)
1978: Prematurity and perinatal mortality in the rhesus
({\it Macaca mulatta}): Relationship to birth weight and
gestational age. Biology of the Neonate 34,129-145.

\qparr
Shaukat (S.S.), Siddiqui (Z.S.), Aziz (S.) 1999: Seed size
and its effects on germination growth and seedling survival
in {\it Acacia nilotica} subspecies {\it indica}.
Pakistan Journal of Botany 31,2,253-263.

\qparr
Smith (M.A.) 2011: Effect of nicotine on {\it Caenorhabditis elegans}
survival, reproduction, and gene expressions.
Development of an invertebrate animal model for drugs abuse.
Master thesis, University of East Carolina.

\qparr
Sterk (A.A.) 1975: Demographic studies of {\it Anthyllis vulneraria}
in the Netherlands. Acta Botanica Neerlandica  24,315-337.

\qparr
Tew (J.C.) 2008: The nematode {\it Caenorhabditis elegans}, 
a model organism for
the study of methyl mercury toxicity.
Master thesis, Michigan State University.

\qparr
US Department of Agriculture 1992: Preweaning morbidity and
morality. National Swine Survey. Fort Collins, Colorado.\qL
[The survey was conducted over a one-year period 
from December 1989 to January 1991. It relied on estimates
from a sample which represented 95\% of the US swine population.]

\qparr
Valen (L. Van) 1975: Life, death, and energy of a tree.
Biotropica 7,4,259-269.\qL
[This is one of the few articles that we could find whose
data are suitable for studying infant mortality in plants.
Another feature which makes it commendable is the
fact that the author explains in detail and very clearly how he
collected the data.]

\qparr
Wu (D.), Cypser (J.R.), Yashin (A.I.), Johnson (T.E.) 2008:
The U-shaped response of initial mortality in 
{\it Caenorhabditis elegans} to mold heat shock.
The Journals of Gerontology. Series A: Biological
Sciences and Medical Sciences 63,7,660-668.

\qparr
Zoological Society of London Archives: Daily occurrences.
[This is an archive source which consists in unpublished handwritten data.]

\end{document}